\documentclass{aa}
\usepackage[utf8]{inputenc}
\usepackage[varg]{txfonts}
\usepackage{bm}
\usepackage{mathtools}
\usepackage{amssymb}
\usepackage{graphicx}
\usepackage{epstopdf}
\usepackage{natbib}
\usepackage{color}
\usepackage[colorlinks=true,     linkcolor=blue, citecolor=blue, filecolor=blue, urlcolor=blue]{hyperref} 

\bibpunct{(}{)}{;}{a}{}{,}
\pdfoutput=1
\usepackage{multirow}
\usepackage{booktabs}

\newlength{\verticalcompensationlength}
\newcounter{verticalcompensationrows}

\begin{document}

   \title{Disruption of exo-asteroids around white dwarfs and the release of dust particles in debris rings in co-orbital motion}

   \author{Kyriaki I. Antoniadou\inst{\ref{inst1},\ref{inst2}} \and   Dimitri Veras\inst{\ref{inst3},\ref{inst4},\ref{inst5}}
          }

   \institute{Department of Mathematics ``Tullio Levi-Civita'', University of Padua, 35121, Padua, Italy 
   \\ \email{kyriaki.antoniadou@unipd.it}\label{inst1} 
   \and
        Department of Physics, Aristotle University of Thessaloniki, 54124, Thessaloniki, Greece \label{inst2}
   \and
   % Department of Physics, Democritus University of Thrace, 65404, Kavala, Greece \label{inst3}
 %  \and
    Centre for Exoplanets and Habitability, University of Warwick, Coventry CV4 7AL, United Kingdom \label{inst3}
   \and
    Centre for Space Domain Awareness, University of Warwick, Coventry CV4 7AL, United Kingdom\label{inst4}
   \and
        Department of Physics, University of Warwick, Coventry CV4 7AL, United Kingdom\label{inst5}
         }

\titlerunning{Co-orbital dynamics in rings}

\authorrunning{K. I. Antoniadou and D. Veras}

\date{Received XXXX; Accepted YYYY}

  \abstract
  % context heading (optional), leave it empty if necessary
   {Close to the Roche radius of a white dwarf (WD), an asteroid on a circular orbit sheds material that then adopts a very similar orbit. Observations of the resulting debris show a periodic behaviour and changes in flux on short timescales, implying ongoing dynamical activity. Additional encounters from other minor planets may then yield co-orbital rings of debris at different inclinations. The structure, dynamics, and lifetime of these debris discs remains highly uncertain, but is important for understanding WD planetary systems.}   
   {We aim to identify and quantify the locations of co-orbitals in WD--asteroid--dust particle three-body systems by exploring the influence of 1:1 resonant periodic orbits. We begin this exploration with co-planar and inclined orbits in the circular restricted three-body problem (CRTBP) and model the dynamical evolution of these exosystems over observable timescales. The mass ratio parameter for this class of systems ($\approx 2 \times 10^{-11}$) is one of the lowest ever explored in this dynamical configuration.}
   {We computed the periodic orbits, deduced their linear stability, and suitably seeded the dynamical stability (DS) maps. We carried out a limited suite of $N$-body simulations to provide direct comparisons with the DS maps.}
   {We derive novel results for this extreme mass ratio in the CRTBP, including new unstable 3D families. We illustrate through the maps and $N$-body simulations where dust can exist in a stable configuration over observable timescales across a wide expanse of parameter space in the absence of strong external forces.  }
   {Over a timescale of 10 years, the maximum orbital period deviations of stable debris due to the co-orbital perturbations of the asteroid is about a few seconds. Unstable debris in a close encounter with the asteroid typically deviates from the co-orbital configuration by more than about 20~km and is on a near-circular orbit with an eccentricity lower than $\approx$0.01.}

   \maketitle
\nolinenumbers
%-------------------------------------------------------------------
\section{Introduction} \label{intro}

Transiting rocky debris is a signpost of the most dynamically active white dwarf (WD) planetary systems \citep{vander15,vander20,vander21,guid21,farihi22}. Unlike exoplanet transits around main-sequence stars, which feature a characteristic single solid-body dip in the light curve, minor planets that break up around WDs contain asymmetrical, sharp, shallow, sometimes periodic and sometimes ephemeral transit features on a nightly basis \citep{gan16,rapp16,gary17,izq18,rap18,agd24}.

Although the transit observations are too complex to be explained by a simple disruption model, analytical progress has been made. For WD 1145+017, the first WD discovered with transiting debris \citep{vander15}, signatures corresponding to small periodic dips in the transit curves are assumed to be broken-off fragments and have been leveraged to estimate both the mass ($\approx 10^{20}$~kg) and orbital eccentricity ($\approx 0$) of the progenitor asteroid \citep{rapp16,gurr17}. Furthermore, the frequency and shape of the transit features themselves have been linked to density, composition, and layering of the progenitor asteroid \citep{ver17,duvv20}.

More generally, the shedding of mass that follows the progenitor orbit has now been investigated extensively \citep{deb12,ver14,mal20a,mal20b,Li2021,brou22,brou23}, as has the process of circularising the debris \citep{vleg15,nix20,oco20,mal21,vbz22}. In contrast, the analytical structure of this phase space has remained relatively unexplored, despite its importance \citep{vmtg16}. Uncovering the stable and unstable co-orbital regions of these systems can aid in future modelling efforts, and it may help to explain the observations.

Dynamically speaking, WD planetary systems provide extreme examples of the circular restricted three-body problem (CRTBP). The masses of WDs are about half the mass of the Sun, and an orbiting minor planet represents the secondary, leading to extremely low mass ratios. Furthermore, the Roche radius of a WD is located at about $1R_{\odot}$. Hence, the secondary orbits are on the scale of many hours, and spatial distances on the scale of $10^{-3}$~au become important (similar e.g. to Saturn's rings).

The uncertainty about the lifetimes of these debris discs and rings adds to the complexity of these systems \citep{gir12,verhen20}. If both discs are sufficiently long lived and the flux of asteroids into the WD Roche radius is sufficiently high, then we expect the debris to be replenished anisotropically. Simulations have demonstrated that asteroids (and potentially debris) would enter the Roche radius at a wide variety of inclinations \citep{vg21}. Furthermore, observations do not yet provide a clear picture of the three-dimensional shapes of these debris discs \citep{man16,man21,ball22,Goksu2024}. These arguments emphasise the importance of investigating the inclined CRTBP (or 3D-CRTBP) in addition to the co-planar CRTBP (or 2D-CRTBP).

The paper is organised as follows. In Sect. \ref{coorb}, a brief literature review of co-orbital motion is provided. In Sect. \ref{main}, the main aspects of the method are presented, that is, the definition of periodic orbits and mean motion resonances (MMRs), and the linear stability and long-term evolution of systems hosted in specific regions in phase space. In Sect. \ref{cfams}, we compute the 2D and 3D families of 1:1 resonant periodic orbits in the CRTBP. In Sect. \ref{regs}, we provide an extended view of the phase space around the WD--asteroid--dust particle system via dynamical stability (DS) maps, which reveal all types of orbits in co-orbital dynamics. In Sects. \ref{nbody} and \ref{sims}, we describe the setup and results of the $N$-body simulations. We discuss our results in Sect. \ref{dis}, and we finally conclude in Sect. \ref{con}.

\section{Brief history and aspects of co-orbital motion} \label{coorb}
\citet{brown11} and \citet{jackson1913} were the first to clearly identify families of what today is referred to as quasi-satellite (QS) orbits in the three-body problem (TBP), based on some preliminary results on retrograde periodic orbits treated by \citet{darwin1897}, which were then called retrograde satellite orbits. In general, this motion takes place outside of the Hill sphere surrounding the minor primary body of the restricted TBP (RTBP), and these trajectories are much closer to it than to the major primary body \citep{Hill1878,kogan88,kogan90,Mikkola97}. These orbits are not related with the tadpole (TP) or horseshoe (HS) orbits and can exist far from the Lagrangian points $L_3$, $L_4$, and $L_5$ \citep[see e.g.][for the behaviour of the resonant phase around these different types of orbits]{Christou00a,Nesvorny02,Mikkola06}. 

For clarity, we refer to co-orbital motion as two bodies that share the same orbit, and QS orbits are a special class of TBP orbits that belong to the family of stable symmetric periodic orbits in 1:1 MMR. In the planar circular RTBP (2D-CRTBP), this family of stable periodic orbits was called family $f$ by \citet{jackson1913,Broucke68,Henon69,HenonGuyot70,Benest74,Bruno94}, and \citet{Pousse17}. The orbits of family $f$ are generated by family $E_{11}^+$ \citep{hen97}, when the problem mass parameter $\mu>0$, and they terminate at a collision orbit with the major primary body \citep{Henon69}. In H\'enon's notation, the generating orbit of family $E_{11}^+$ is a third-species orbit (the orbits of the minor primary body and the mass-less body coincide when $\mu\rightarrow 0$). The QS orbits of family $f$ are stable for $\mu<0.0477$ \citep{HenonGuyot70,Benest74}. 

In the 2D-CRTBP, there exist Lyapounov families with unstable symmetric periodic orbits and stable asymmetric periodic orbits. The former family, called family $b$ by \citet{hen97}, consists of HS orbits that are generated by family $E_{11}^-$ in H\'enon's notation when $\mu>0$, and it includes $L_3$ (the orbits of the minor primary body and the mass-less body are diametrically opposed when $\mu\rightarrow 0$). The latter family, called family $E_{11}^a$ by H\'enon, includes the short-period\footnote{The period of short-period orbits tends to $2\pi$, whereas the period of the long-period orbits tends to infinity when $\mu\rightarrow 0$.} orbits emanating from $L_4$ and intersects family $E_{11}^-$ when the linear stability changes (see Sect. \ref{cfams} herein). This family has a mirror image that emanates from $L_5$. \citet{Pousse17} called these families of periodic orbits in the rotating frame of the 2D-CRTBP $\mathcal{L}_3$ and short-period $\mathcal{L}_4^s$ (or $\mathcal{L}_5^s$). We do not discuss the family of long-period unstable asymmetric periodic orbits, which terminates in the family of short-period asymmetric ones \citep[see e.g.][]{deprit67}, and the families starting from $L_1$ (prograde satellite motion in family $c$) and $L_2$ (family $a$) \citep[see e.g.][]{Henon69,hen97}. 

\begin{figure}\centering
\includegraphics[width=0.5\textwidth]{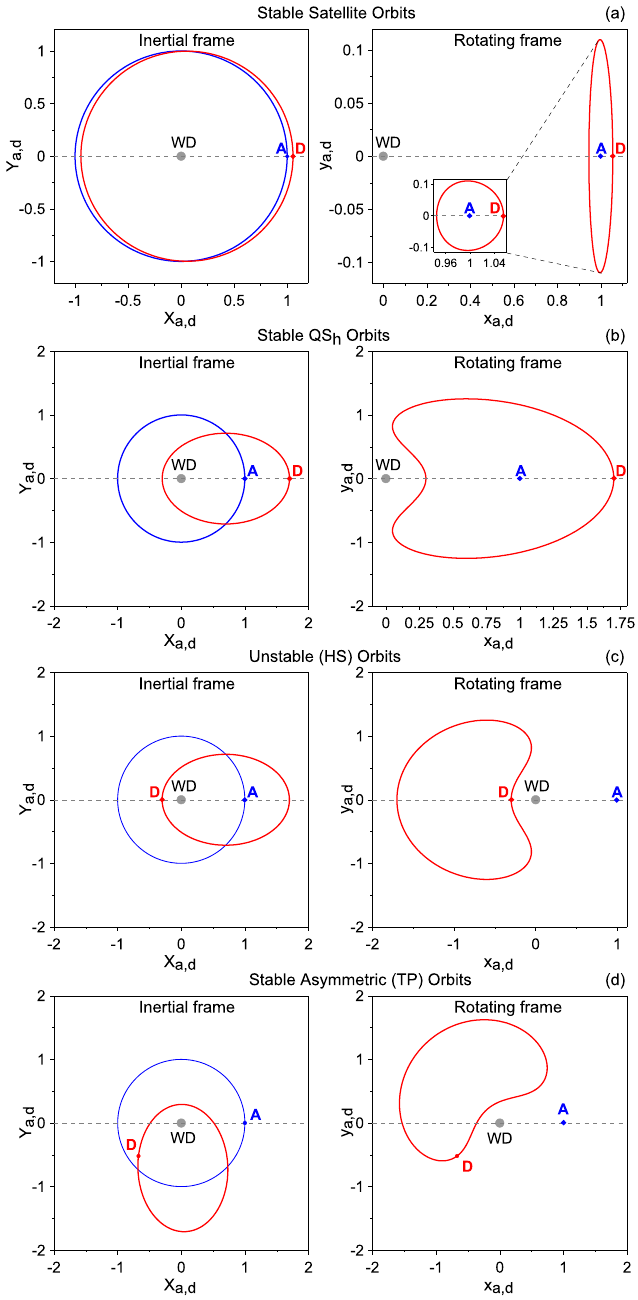}
\caption{Periodic orbits in the 2D-CRTBP, where the asteroid, denoted by $A$ (in blue), moves on circular orbits ($e_{\rm a}=0$), and the dust particle, $D$ (in red), moves on Keplerian ellipses that rotate around the WD (grey dot), when viewed on an inertial frame of reference (left column). These orbits are exactly periodic in the rotating frame of reference (right column). The orbits were computed for one period, $T=2\pi$, and both bodies orbit the WD with the same period (corresponding in physical space to a distance unit of about $1R_{\odot}$ and to an orbital period of about 4.5 hours). The red dots represent the position of the dust particle at $t=0$. In panel a, the stable retrograde satellite orbits that belong to family $f$ are demonstrated, where the dust particle moves on almost circular orbit. In panels b and c, the symmetric stable QS$_h$ orbits of family $f$, and the symmetric unstable horseshoe (HS) orbits are shown, respectively. In panel d, the stable asymmetric (tadpole, TP) periodic orbits are shown. The eccentricity of the dust particle chosen for these panels was $e_{\rm d}=0.7$, and the initial conditions were taken from the families presented in Fig.~\ref{fig5}.}
        \label{QSorbits}
\end{figure}

We instead explore four distinct types of orbits. In our case study, namely WD--asteroid--dust particle, we have the categories listed below. 
\begin{itemize}
        \item When the motion of the dust particle takes place close to the asteroid and the gravitational perturbation of the WD is assumed to be negligible, the reference system is 'asteroid-centred' and the orbits are almost those of a Keplerian retrograde satellite. The dust particle forms a close binary with the asteroid, and their centre of mass revolves around the WD (Fig.~\ref{QSorbits}a). 
        \item When the motion of the dust particle is quite distant from the asteroid and the gravitational attraction of the WD dominates the orbits, we may use a 'WD-centred' system and compute Keplerian planetary-type orbits (orbits of second kind according to H\'enon). We may have three possible configurations:\\
        (a) The stable configuration of QS$_h$ symmetric orbits is obtained when the bodies are anti-aligned (apsidal difference equal to $180^{\circ}$; Fig.~\ref{QSorbits}b).\\
        (b)  The unstable configuration of HS symmetric orbits arises when the bodies are aligned (apsidal difference equal to $0^{\circ}$) and move in opposite phases (Fig.~\ref{QSorbits}c).\\
        (c) The stable configuration of TP asymmetric orbits is obtained when the bodies are neither aligned nor anti-aligned (Fig.~\ref{QSorbits}d).\\
        The QS$_h$ orbits and the retrograde satellite orbits are separated by the binary QS$_b$ orbits, which belong to a domain in which none of the primaries influences the mass-less body \citep[see e.g.][]{Pousse17,va18}.
          \end{itemize}

Osculating orbital elements can be used to describe these orbits, that is, $a_i$ (semi-major axis), $e_i$ (eccentricity), $i_i$ (inclination), $\omega_i$ (argument of pericentre), $M_i$ (mean anomaly), and $\Omega_i$ (longitude of the ascending node). We also use the notation $\varpi_i=\omega_i+\Omega_i$ for the longitude of pericentre, $\Delta\varpi$ for the apsidal difference, $\Delta\Omega$ for the nodal difference, and $\lambda_i=\varpi_i+M_i$ for the mean longitude. Subscript $i={\rm a}$ denotes the asteroid, and $i={\rm d}$ signifies the dust particle.
In the CRTBP, these orbits render a system consisting of a dust particle and an asteroid moving around the WD with the same orbital period in elliptic, $e_{\rm d}>0$, and in circular, $e_{\rm a}=0$, orbits, respectively. 

In the Solar System, QS orbits have been computed numerically and analytically for the Mars-Phobos system \citep{Lidov94}, while the first results were computed for the Phobos spacecraft mission \citep{kopho88,SZ89}. Stable QS orbits around giant planets in our solar neighbourhood were identified by analytical, semi-analytical, or numerical methods \citep[see e.g.][]{Mikkola97,Namouni99,nachrimu99,Wiegert00,Christou00a,Christou00b,Nesvorny02,Mikkola06,Sidorenko14,Pousse17}. Distant retrograde orbits are particularly informative regarding the space mission design to the Moon, Europa, and other satellites and asteroids \citep[see e.g.][]{Minghu14,Perozzi17,kimura19,pires20,li22}. Asteroid 2003 ${\rm YN_{107}}$ was the first quasi-satellite of Earth that was discovered \citep{Connors04}, followed by studies of others in 1:1 resonance with Earth, such as asteroid 2002 ${\rm AA_{29}}$ \citep{wajer09}, 2004 ${\rm GU_9}$ and 2006 ${\rm FV_{35}}$ \citep{wajer10}, 2013 ${\rm LX_{28}}$ \citep{Connors14}, and 2014 ${\rm OL_{339}}$ and 2016 ${\rm HO_3}$ \citep[][respectively]{fufu14,fufu16}. Other studies involve QS orbits around Venus \citep[e.g.][]{Mikkola04}, Jupiter \citep[e.g.][]{kina07,Namouni18}, and the main asteroid belt co-orbitals \citep[e.g.][]{Christou00b}. 

\citet{madeira24} studied the survival of dust particles in the Didymos-Dimorphos system and their connection with stable QS orbits, while \citet{korte13} explored the QS dust trapping and dynamical evolution in Earth's regime, and \citet{wimu11} modelled the evolution of ring particles in the HS regime in the Janus-Epimetheus system. The Yarkovsky-O'Keefe-Radzievskii-Paddack (YORP) effect on a particular class of Earth and Mars co-orbitals was studied for instance by \citet{fufu18,fufu21}.  

The detectability of extrasolar trojans and bodies on QS-like orbits is in general very difficult because they can be discarded as false positives \citep{laugh02,leleu19}. Nevertheless, this has not discouraged theoretical studies \citep{ray2023a,ray2023b}. 

For reasons of completeness, we refer to the computation of families of QS orbits in the planar general TBP (2D-GTBP). When perturbation was added to the problem and a continuation with respect to the mass of the mass-less body was performed \citep{hadj75}, the QS orbits of family $f$ generated stable symmetric periodic orbits in the 1:1 MMR in the so-called family $S$ in the 2D-GTBP computed for various mass ratios by \citet{hpv09} and \citet{hv11b}. Family $S$ consists of planetary-type (computed in the heliocentric system) and satellite-type orbits (computed in the planetocentric system). This group of families $S$ was later called $g(f_1,E_a)$ by \citet{va18}. Additionally, numerical, analytical, and semi-analytical treatments of the phase space were conducted for example by \citet{Schwarz09,Giuppone10,Robutel13,Leleu17,leleu18,nied20,cout22,sid24} because planetary configurations in which co-orbital motion could be exhibited are very intriguing \citep[see e.g.][]{Giuppone12,Funk13,Lillo18,balsa23}.

\section{Method}\label{main}
\subsection{Notion of periodic orbits related to MMRs and resonant angles}
We aim to determine the location of QS, HS, and TP orbits in the 1:1 MMR and explore the influence of the 1:1 resonant symmetric and asymmetric co-planar or inclined periodic orbits on the dynamical behaviour of dust particles sharing the same orbit with an exo-asteroid within a debris disc or ring around a WD. In general, the MMR acts as a phase-protection mechanism safeguarding any planetary system even when the orbits are highly eccentric \citep{av16}. 

The periodic orbits indicate the exact location of the MMR in phase space and are fundamental for understanding the resonant dynamics of each problem. They coincide with the fixed or periodic points on a Poincar\'e surface of section and with the stationary equilibrium  points of an appropriately averaged Hamiltonian, as long as the latter is sufficiently accurate \citep{hadj93as}. The periodic orbits are not isolated in phase space. They are continued mono-parametrically (see Sect. \ref{setup33}) and form characteristic curves or families \citep{hadj75,hen97}. These families of periodic orbits may be either generated by bifurcation points, or are isolated.% \citep[see e.g.][]{kiaasl}.

We defined the resonant angle $\theta=\lambda_{\rm a}-\lambda_{\rm d}$ \citep{murray} in order to distinguish the various families of periodic orbits in the 2D- and the 3D-RTBPs that dominate and shape each region in phase space. The value of the resonant angle remains constant when computed on the exact periodic orbit. In the following, we discuss the  behaviour of the angles (libration or rotation) that is linked with the linear stability of the periodic orbits and the vicinity of the latter in phase space.  

\begin{figure}\centering
\includegraphics[width=0.5\textwidth]{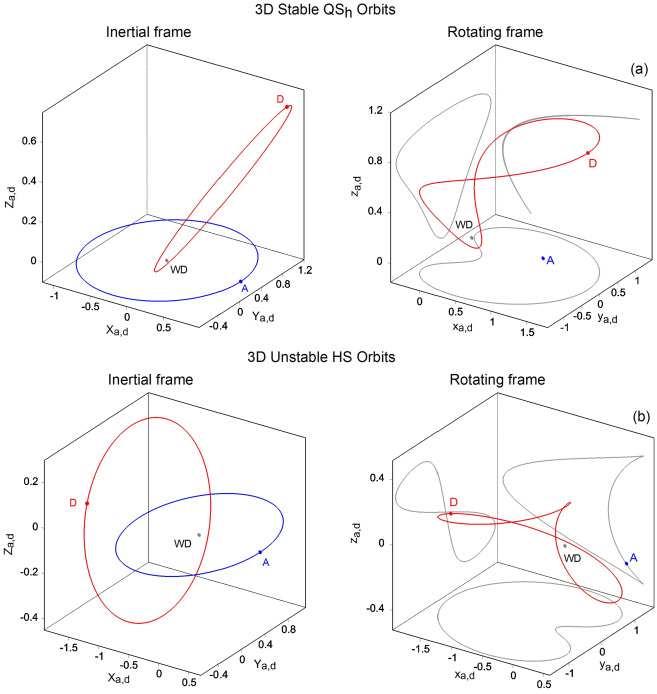}
\caption{Stable (QS$_h$) and unstable (HS) inclined periodic orbits in the 3D-CRTBP with $i_{\rm d}=30^{\circ}$ (panel a) and $i_{\rm d}=50^{\circ}$ (panel b). These orbits belong to the 3D families $F^{I}$ with periodic orbits symmetric with respect to the $xz$-plane and $G^{II'}$ (with periodic orbits symmetric with respect to the $x$-axis, both families are shown Fig. \ref{fig6}). These 3D families may either bifurcate from the 2D family $I$ (or $I'$) or from the 2D family $II$ (or $II'$) shown in Figs. \ref{fig3}-\ref{fig4}). Presentation as in Fig.~\ref{QSorbits}.}
        \label{QSorbits3D}
\end{figure}

\subsection{Linear stability and long-term planetary stability}
We computed the linear stability \citep{marchal90} of the periodic orbits by calculating the eigenvalues of the monodromy matrix of the linearised equations of motion, whose number depends on the degrees of freedom of each problem. If, and only if, all the eigenvalues lie on the unit circle, the periodic orbit is classified as linearly stable. Linearly stable periodic orbits, either planar or spatial, constitute the backbone of stability domains in phase space, where invariant tori exist and the motion is regular and bounded. Therefore, any planetary system therein will survive for long-time spans. In these regions, the resonant angles and the apsidal difference librate about $0^{\circ}$ or $180^{\circ}$ when the periodic orbit is symmetric, that is, when it is invariant under the fundamental symmetry $\Sigma: (t,x,y) \rightarrow (-t,x,-y)$ \citep[see][]{hen}, or about other values when it is asymmetric and $\Sigma$ maps it to its mirror image \citep{voyhadj05}.

When the periodic orbits are linearly unstable, chaos manifests itself and leads to irregular evolution. This may not change the configuration of the system significantly when the chaos is weak. When the chaos is strong, planetary disruption may occur and lead to close encounters, collisions, or escapes, and hence, to planetary destabilsation. In the vicinities of these chaotic seas, the resonant angles and the apsidal difference rotate. 

We also took an intrinsic property of the planar periodic orbits into account, namely the vertical stability, by computing the vertical stability index \citep{hen}. A system of celestial bodies moving on different inclined planes can evolve stably when their mutual inclinations are low and when they are located in the neighbourhood of vertically stable 2D periodic orbits. The orbits that are vertically critical (vco) act as bifurcation points that generate 3D periodic orbits. Celestial bodies with a high mutual inclination can only survive when they are located in the neighbourhood of linearly stable 3D resonant periodic orbits \citep[see e.g.][]{av14}.

Therefore, the methodical computation of families of periodic orbits and the deduction of their linear stability are precise and rigorous diagnostic tools to guide the exploration of the phase space. In this study, we consider both symmetric and asymmetric  1:1 resonant periodic orbits. Planar stable asymmetric orbits were for example found in the 2D-GTBP by \citet{Giuppone10,hv11b}. Families of 2D asymmetric periodic orbits were also found to be vertically stable in the 2D-GTBP by \citet{avv14}, and hence, these co-orbital bodies will survive small deviations from co-planar motion.

\subsection{System set-up}\label{setup33}
We first considered a WD, an asteroid, and a dust particle as point masses, $m_{\rm WD}$, $m_{\rm a}$, and $m_{\rm d}=0$, respectively, and defined the problem parameter as $\mu=\frac{m_{\rm a}}{m_{\rm WD}+m_{\rm a}}$. In this configuration, the motion of the dust particle does not affect the motion of the WD and the asteroid while it moves under their gravitational attraction. In 1:1 resonance, the semi-major axes of the asteroid and the dust particle are almost equal ($a_{\rm a} \approx a_{\rm d}$).

In the inertial frames of reference, for instance, $OXY$ or $GXYZ$, the periodic orbits correspond to almost Keplerian ellipses, which are described by the osculating elements and rotate around the major primary body. In Fig.~\ref{QSorbits}, the Keplerian ellipses presented in the 2D inertial frame (left column) are computed for one period, that is, $T=2\pi$, and hence, their rotation around the WD is not apparent. 

We introduced suitable rotating frames of reference to reduce the degrees of freedom of the systems and define the periodic orbits. In the 2D-RTBPs, the rotating frame of reference, for instance, $Oxy$, is centred at the centre of mass of the primaries, that is, the WD (major primary body) and the asteroid  (minor primary body), while the motion of the asteroid is restricted on the $Ox$-axis \citep[see e.g.][]{hach75}. The dust particle describes a periodic motion on the $Oxy$-plane (right column of Fig.~\ref{QSorbits}). The exact resonance is defined as the 1:1 resonant periodic motion in this frame.

In the 3D-RTBPs, the dust particle was allowed to evolve on inclined orbits with respect to the orbital plane of the WD and the asteroid (see e.g. the left column in Fig. \ref{QSorbits3D}). The origin of the 3D rotating frame, $Gxyz$, called $G$, coincides with the centre of mass of the primaries, the $Gz$-axis is perpendicular to the $Gxy$-plane, and its $Gx$-axis is directed from the WD to the asteroid \citep[see e.g.][for more details on various aspects of the RTBPs]{sze,murray}. 

An orbit, \textbf{Q}(t), is considered as periodic when it satisfies the condition $\textbf{Q}(0) = \textbf{Q}(T)$, where \textbf{Q}(0) is the set of initial conditions (positions and velocities defined in the rotating frame) at t=0, and $T$ is the orbital period. We assumed a solution $\textbf{Q}(t)=(x_{\rm d}(t),x_{\rm a}(t),y_{\rm d}(t),\dot x_{\rm a}(t),\dot x_{\rm d}(t), \dot y_{\rm d}(t))$ in the 2D-RTBP.  This solution is  periodic with a period $T$ when it satisfies the periodic conditions
\begin{equation} \label{EqPeriodCond}
\begin{array}{ll}
\dot x_{\rm a}(T)=\dot x_{\rm a}(0)=0, & \\
x_{\rm a}(T)=x_{\rm a}(0), & \\ 
x_{\rm d}(T)=x_{\rm d}(0) ,& y_{\rm d}(T)=y_{\rm d}(0), \\
\dot x_{\rm d}(T)=\dot x_{\rm d}(0), & \dot y_{\rm d}(T)=\dot y_{\rm d}(0). 
\end{array}
\end{equation} 
Given the Poincar\'e surface of section at $y_{\rm d}=0$ and the fact that  $x_{\rm a}=1-\frac{m_{\rm a}}{m_{\rm WD}+m_{\rm a}}$ is constant and defined by the normalisation we used in the CRTBP, a symmetric periodic orbit that perpendicularly crosses the $Ox$-axis ($\dot x_{\rm d}(0)=0$) is defined as a point on the plane of initial conditions $\{(x_{\rm d}(0),\dot y_{\rm d}(0))\}$. The asymmetric periodic orbit can for example be defined in the space $\{(x_{\rm d}(0), y_{\rm d}(0), \dot x_{\rm d}(0), \dot{y}_{\rm d}(0))\}$ when $\dot x_{\rm a}(0)= 0 $. 

For the 3D periodic orbits shown in the right column of Fig. \ref{QSorbits3D}, the $xz$-symmetric periodic orbits (top panel) can be represented by a point in the 3D space of the initial conditions
$\{(x_{\rm d}(0),z_{\rm d}(0),\dot y_{\rm d}(0))\}$ and the $x$-symmetric ones (bottom panel) by $\{(x_{\rm d}(0),\dot y_{\rm d}(0),\dot z_{\rm d}(0))\}$.

The families are formed during the mono-parametric continuation when a parameter, for example, $z_{\rm d}$, is changed for the $xz$-symmetry, and the rest are differentially corrected in order to satisfy the respective periodicity condition. Continuation with respect to the mass of a body is also possible for example by varying the mass of $m_{\rm a}$ and keeping $z_{\rm d}$ fixed, for instance, while the respective periodicity condition still holds. The limitations of this continuation method can be found in \citet{hadj75,av12}. The periodicity conditions that the periodic orbits must fulfil were described in greater detail for instance in \citet{kiaasl} for the 2D-RTBPs and in \citet{spa} for the 3D-RTBPs.

To numerically computate the periodic orbits we considered, we located without loss of generality the asteroid at $a_{\rm a}=1$, and its period was therefore $T_0=2\pi$. We also normalised the masses and the gravitational constant, $G$, so that the sum of the former was equal to unity, namely $m_{\rm WD}+m_{\rm a}+m_{\rm d}=1$ or $m_{\rm WD}=1-m_{\rm a}$ and $G=1$.

In physical units, given the WD 1145+017 system, we adopted a WD mass $m_{\rm WD} = 0.6m_{\odot}$. Then, we assumed an asteroid with a mass equal to approximately $1:10^{\rm th}$ the mass of Ceres, for instance, $2.5 \times 10^{19}$ kg, located at $a_{\rm a}=0.0054$ au around the WD \citep[see e.g. Table 1 in][]{rapp16}. In this configuration, we derived a $\mu$ equal to $2.1186\times 10^{-11}$. To our knowledge, this value is one of the lowest ever studied for the 2D and 3D CRTBP. A comparably low value of $\mu$ ($10^{-9}$) for the numerical computation of QS and HS orbits in the CRTBP was adopted by \citet{Lidov94} for the specific case of retrograde orbits in the 2D-CRTBP and by \citet{llibre01} for the Saturn-Janus-Epimetheus system, respectively. \citet{Pousse17} explored the domains around QS orbits for $\mu\geq 10^{-7}$. Finally, numerical computations yielding the extent of regular domains around HS and TP orbits in 1:1 MMR as a function of $\mu$ were performed by \citet{liwi20}.

\subsection{Visualisation of the phase space}
In addition to the linear stability of the periodic orbits, which provides a direct insight into the long-term stability of any system in their vicinity, we also computed DS maps with the use of a chaotic indicator to visualise the extent of each domain. We adopted a version of the FLI, called detrended fast Lyapunov indicator (DFLI), that was established as reliable and accurate by \citet{voyatzis08}. We constructed $200 \times 100$ grid planes and chose a maximum integration time for the computation of the DFLI equal to $t_{\rm max}=2.5$ Myr, that is, almost 4.8 billion orbits of the dust particle, which was deemed adequate for these systems. We halted the integrations either when DFLI$(t)>30$ or when $t_{\rm max}$ was reached. Orbits with ${\rm DFLI}<2$ were classified as stable. Extensive numerical computations have shown that a system can be calssified as chaotic when ${\rm DFLI}>15$, as the DFLI continues to increase steeply thereafter \citep{numan2014}. %In order to highlight the structure of the phase space, we used 16, instead of 30, as the maximum value incorporated in colour palette.

\section{Families of periodic orbits in the CRTBP}\label{cfams}

In this model for the 1:1 MMR, the asteroid moves on a circular orbit $(e_{\rm a}=0)$ with $a_{\rm a}=1$, and the dust particle moves on an eccentric orbit $(e_{\rm d}>0)$ with $a_{\rm d}\approx 1$. In the following sections, we present the families of 2D and 3D periodic orbits in the CRTBP. 

\subsection{2D-CRTBP}

\begin{figure}
\includegraphics[width=0.45\textwidth]{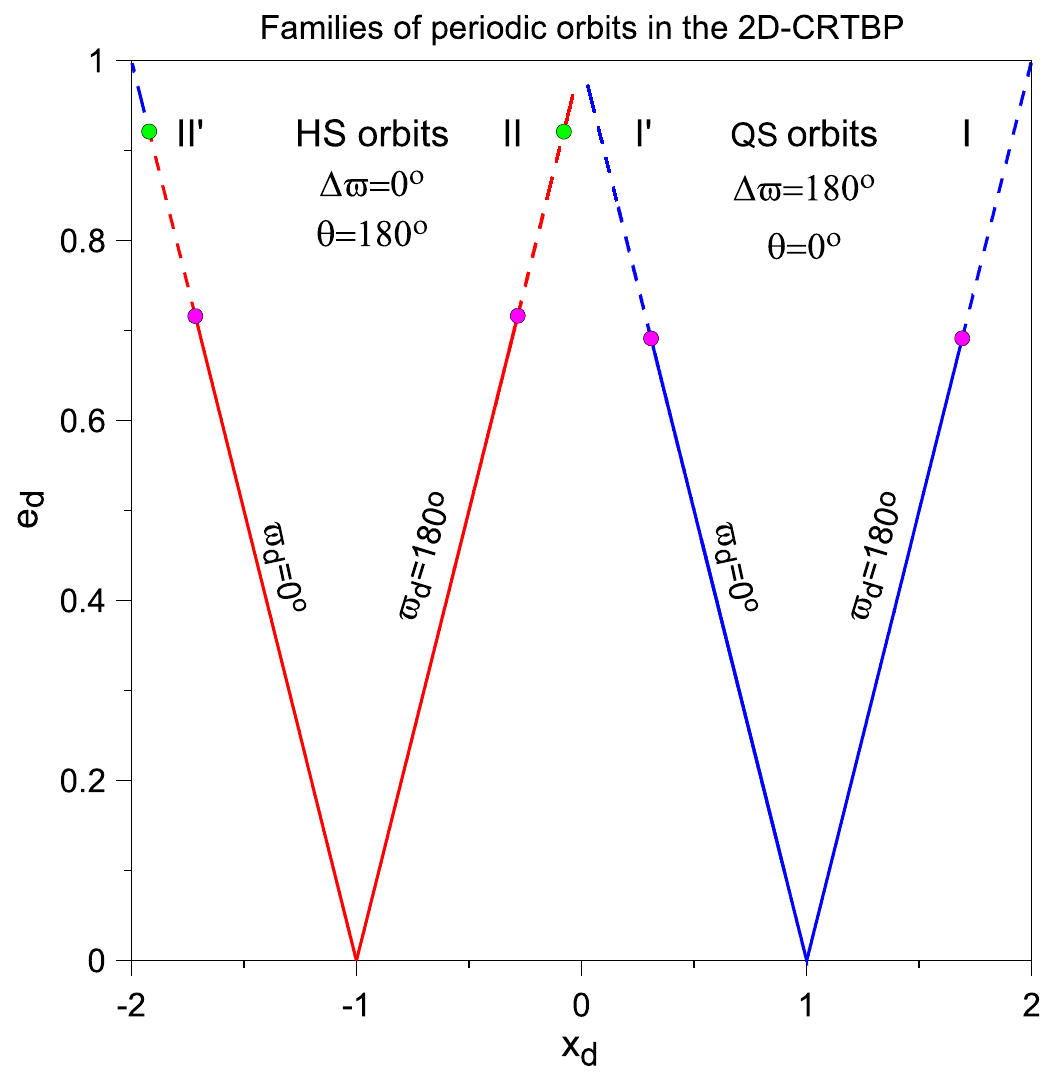}
\caption{Families of QS and HS orbits in the 2D-CRTBP presented on the $(x_{\rm d},e_{\rm d})$ plane, where $x_{\rm d}$ is the position of the periodic orbit (dust particle) in the rotating frame of reference for $\mu=2.1186\times 10^{-11}$. Families denoted by $I$ correspond to family $f$, which consists of QS orbits, namely stable (blue) symmetric periodic orbits with $\Delta\varpi=180^{\circ}$, and $II$ denotes the family of HS orbits, i.e. unstable (red) symmetric periodic orbits with $\Delta\varpi=0^{\circ}$. Primed (or unprimed) symbols indicate the direction (positive or negative crossing) of the orbit on the Poincar\'e surface of section at $y_{\rm d}=0$ and $\dot y_{\rm d}>0$ (or $\dot y_{\rm d}<0$), since the two branches of each family are equivalent (see also Table \ref{tab1}). The magenta and green dots correspond to the vcos, and the solid (dashed) lines demonstrate the vertical stability (instability).}
        \label{fig3}
\end{figure}

\begin{table}%[h!]
\centering
\caption{Equivalent configurations of 1:1 resonant periodic orbits in the 2D-CRTBP presented in Fig. \ref{fig3}, namely $I$ with $I'$ and $II$ with $II'$, at $t=0$ and at $t=T/2$.}
\begin{tabular}[b]{lcccc}
\toprule
Family& $\varpi_{\rm a} (^{\circ})$     &$M_{\rm a} (^{\circ})$ &$\varpi_{\rm d} (^{\circ})$  &$M_{\rm d} (^{\circ})$\\%&$\Delta\varpi$&$\theta_1$\\
\cmidrule{1-5}
$I$         &0      &0             &180       &180\\
\cmidrule{1-5}
$I'$                     &180    &180           &0         &0\\
\cmidrule{1-5}
$II$                     &180  &180           &180       &0\\
\cmidrule{1-5}
$II'$       &0     &0             &0         &180\\
\bottomrule
\end{tabular}\label{tab1}
\end{table}

\begin{figure}%[!h]%\begin{figure*}[!h]\centering
\includegraphics[width=0.45\textwidth]{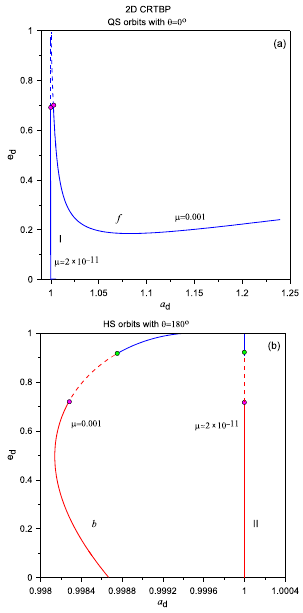}
\caption{Families $I$ of QS orbits (panel a) and $II$ of HS orbits (panel b) on the $(a_{\rm d},e_{\rm d})$ plane for our model ($\mu=2.1186\times 10^{-11}$ presented in Fig.~\ref{fig3}), together with a comparison with $\mu=0.001$ and families $f$ and $b$ to highlight the proximity of this model to the unperturbed case.}
        \label{fig4}
\end{figure}%\end{figure*}

\begin{figure}%[!h]%\begin{figure*}[!h]\centering
\includegraphics[width=0.45\textwidth]{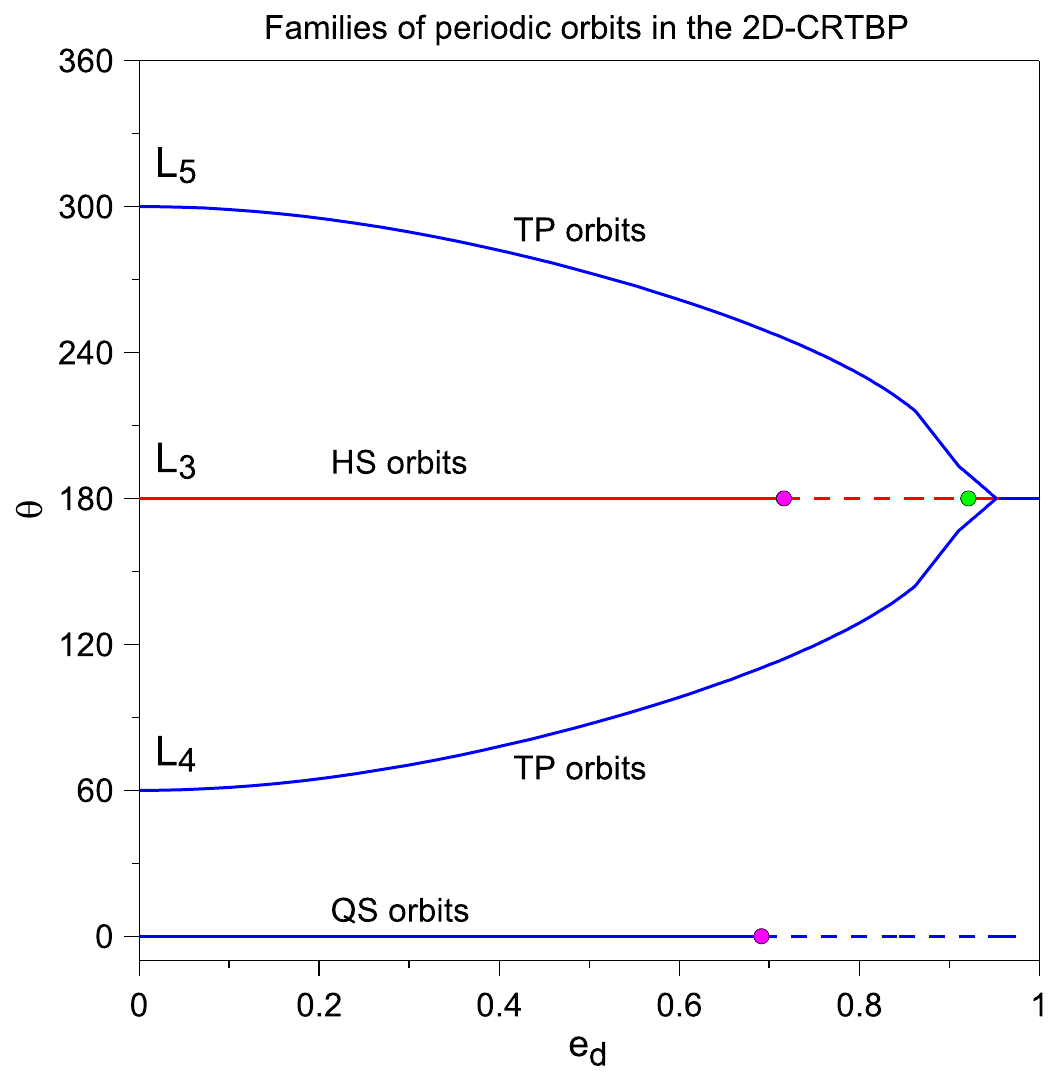}
\caption{Families of QS, HS, and TP orbits on the $(e_{\rm d},\theta)$ plane. Presentation as in Fig. \ref{fig3}.}
        \label{fig5}
\end{figure}%\end{figure*}

In Fig.~\ref{fig3} we present the families of periodic orbits in the 2D-CRTBP on the $(x_{\rm d},e_{\rm d})$ plane, namely the QS and HS orbits. In Fig.~\ref{fig4}, we present the families $I$ (stable branch) and $II$ (unstable branch) on the $(a_{\rm d},e_{\rm d})$ plane and provide a comparison with $\mu=0.001$ to highlight the fact that we approach the unperturbed case ($\mu\rightarrow 0$). Symbol $I$ corresponds to family $f$ which consists of stable (blue coloured) periodic orbits, where the asteroid and the dust particle are in anti-alignment ($\Delta\varpi=180^{\circ}$ and $\theta=0^{\circ}$), while symbol $II$ denotes family $b$ of unstable (red coloured) periodic orbits, where the bodies are in alignment ($\Delta\varpi=0^{\circ}$ and $\theta=180^{\circ}$). We note that the two branches shown for each family are equivalent and differ only in the positive or negative direction (namely $\dot y_{\rm d}>0$ or $\dot y_{\rm d}<0$) of the orbit for which the Poincar\'e surface of section at $y_{\rm d}=0$ was chosen. More particularly, we obtain the configurations shown in Table \ref{tab1}, which are equivalent in pairs, that is, $I$ with $I'$ and $II$ with $II'$, at $t=0$ and at $t=T/2$. We provide them all here because the continuation process in the 3D-CRTBP, 2D- and 3D-ERTBP (which will be shown in a future study) was computationally easier when we started from the same bifurcation point but with a different cross section, since cusps appear in the rotating frame of reference. 

In Fig. \ref{fig5} we show the family of asymmetric periodic orbits, which consists of TP orbits, and starts from $L_4$ at $\theta=60^{\circ}$ and ends at a symmetric periodic orbit of HS type at $\theta=180^{\circ}$, where the linear stability changes. We also present the mirror family, which starts from $L_5$ at $\theta=300^{\circ}$. These families are both horizontally and vertically stable (solid blue lines).

\subsection{3D-CRTBP}
We considered the families of QS, HS, and TP orbits presented in the 2D-CRTBP and computed their vertical stability. Two types of vcos generate 3D periodic orbits with two different symmetries \citep{hen,ichmich80}, namely with respect to the $xz$-plane or the $x$-axis of the rotating frame $Gxyz$. We call the generated families F and G, respectively, and use a superscript to denote the planar family to which the vco (bifurcation point) belongs. We add a hat to represent the respective vcos. For example, vco ${\widehat{F}}^I$ belongs to the 2D family $I$ and generates the 3D family $F^I$ of $xz$-symmetric (magenta vco) periodic orbits. In Table \ref{tab2}, we provide the eccentricity values for each vco. The family of asymmetric periodic orbits (TP orbits) is vertically stable.

\begin{table}[h!]
\centering
\caption{Bifurcation points (vcos) from the 2D-CRTBP to the 3D-CRTBP shown in Fig. \ref{fig3}.}
\begin{tabular}[b]{lccc}
\toprule
vco&    $e_{\rm d}$     &2D family      &3D family\\
\cmidrule{1-4}
${\widehat{F}}^I$   or ${\widehat{F}}^{I'}$     &$0.691$      &$I$ or   $I'$     &${F}^I$  or ${F}^{I'}$   \\
\cmidrule{1-4}
${\widehat{F}}^{II}$ or ${\widehat{F}}^{II'}$   &$0.716$      &$II$ or  $II'$    &${F}^{II}$  or ${F}^{II'}$  \\
\cmidrule{1-4}
${\widehat{G}}^{II}$ or ${\widehat{G}}^{II'}$   &$0.921$      &$II$ or  $II'$    &${G}^{II}$  or ${G}^{II'}$   \\
\bottomrule
\end{tabular}\label{tab2}
\end{table}

The 3D families of unstable orbits, which stem from the unstable family $II'$ (or $II$) of HS orbits in the 2D-CRTBP, have not been presented before for any value of $\mu$. The same holds for the vcos that generate them. For the family $F^I$ of stable QS orbits, we find no significant change in the location of the vco as $\mu\rightarrow 0$, that is, as we approach the unperturbed case ($m_{\rm a}=m_{\rm d}=0$) (see the vcos (magenta dots) in panel a of Fig.~\ref{fig4} and compare e.g. the ${\widehat{F}}^I$ (or ${\widehat{F}}^{I'}$) with the $B_{cs}$ points in Table 1 in \citet{va18}). Nonetheless, we note that as $\mu\rightarrow 0$, the QS domain becomes reachable by low-eccentricity orbits and the trajectories approach the minor primary, that is, the asteroid in our case, while for $\mu=0.001$, this domain exists for $e_{\rm d}>0.18$ (see the top panel of Fig.~\ref{fig4}).

\begin{figure}%[!h]%\begin{figure*}[!h]
\centering
\includegraphics[width=0.4\textwidth]{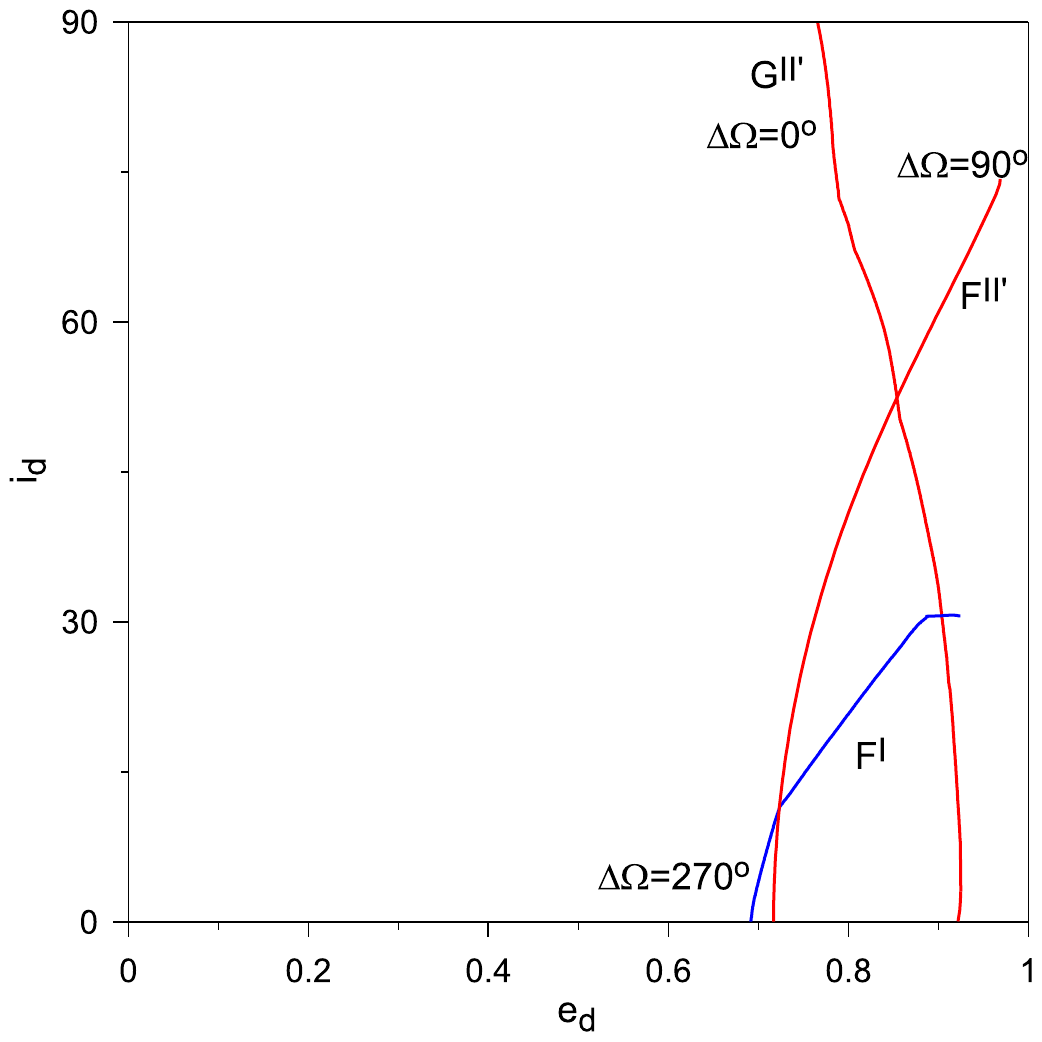}
\caption{Three families, $F^I$, $F^{II'}$, and $G^{II'}$ of 3D periodic orbits in the 3D-CRTBP on the $(e_{\rm d},i_{\rm d})$ plane. The families of QS and HS orbits of the 2D-CRTBP lie on the $x$-axis at $i_{\rm d}=0^{\circ}$. Presentation as in Fig. \ref{fig3}.} 
        \label{fig6}
\end{figure}%\end{figure*}

\begin{figure*}
\includegraphics[width=0.99\textwidth]{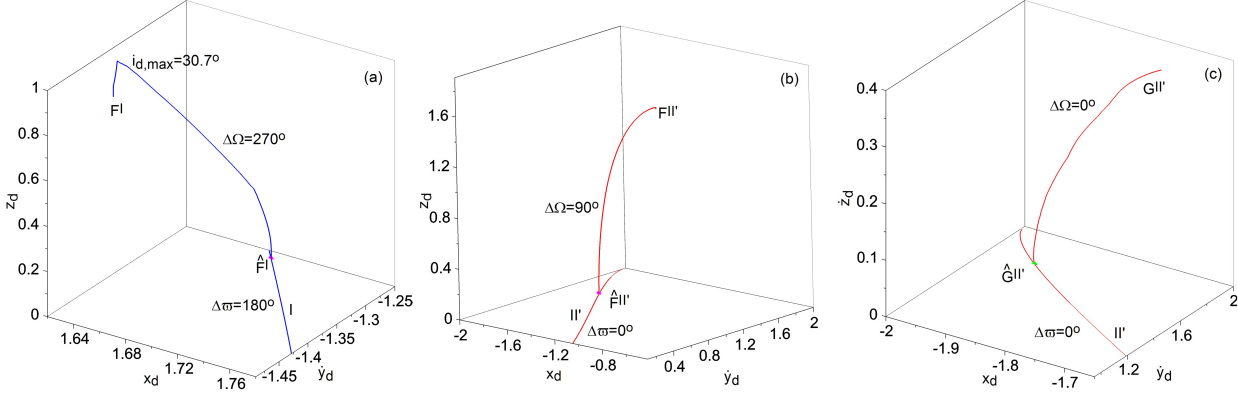}
\caption{Three families of 3D periodic orbits in the 3D-CRTBP of Fig. \ref{fig6} shown in $(x_{\rm d},\dot y_{\rm d}, z_{\rm d})$ space and $(x_{\rm d},\dot y_{\rm d}, \dot z_{\rm d})$ space. The respective families of the 2D-CRTBP together with the bifurcation points (vcos) are also included (at $z_{\rm d}=0$ or $\dot z_{\rm d}=0$). Presentation as in Fig. \ref{fig3}.}   
        \label{fig7}
\end{figure*}

\begin{table}[h!]
\centering
\caption{Configuration of the dust particle for each family in the 3D-CRTBP presented in Fig. \ref{fig6}.}
\begin{tabular}[b]{lccccc}
\toprule
3D family&      $\omega_{\rm d} (^{\circ})$     &$\Omega_{\rm d} (^{\circ})$    &$M_{\rm d} (^{\circ})$ & $\Delta\varpi (^{\circ})$ & $\Delta\Omega (^{\circ})$\\
\cmidrule{1-6}
${F}^I$                         & 270  & 270 & 180 & 180 & 270 \\
\cmidrule{1-6}
${F}^{II'}$   & 270  & 90  & 180 & 0   & 90 \\
\cmidrule{1-6}
${G}^{II'}$   &   0  & 0   & 180 & 0   & 0\\
\bottomrule
\end{tabular}\label{tab3}
\tablefoot{The asteroid $(e_{\rm a}=0)$ has the following angle values: $i_{\rm a}=\omega_{\rm a} = \Omega_{\rm a} = M_{\rm a}=0^{\circ}$ and is located at $a_{\rm a}=1$.}
\end{table}

In Fig.~\ref{fig6} we present the three 3D families of periodic orbits in the 3D-CRTBP on the $(e_{\rm d},i_{\rm d})$ plane. The family $F^I$ consists of stable $xz$-symmetric periodic orbits (QS orbits) up to $i_{d,\rm max}=30.7^{\circ}$. Both $F^{II'}$ and $G^{II'}$ have only unstable periodic orbits (HS orbits) with an $xz$- and $x$-symmetry, respectively. In Table \ref{tab3} we present the configuration of each family. In Fig.~\ref{fig7}, each family is shown in more detail in $(x_{\rm d},\dot y_{\rm d}, z_{\rm d})$ space or $(x_{\rm d},\dot y_{\rm d}, \dot z_{\rm d})$ space, depending on the symmetry of the periodic orbits together with the projections of the planar families and the vcos in the 2D-CRTBP (at $z_{\rm d}=0$ or $\dot z_{\rm d}=0$).

\section{Stable and chaotic QS, HS, and TP orbit regimes in the 3D CRTBP}\label{regs}
We identified the regular domains in phase space on the DS maps by monitoring the libration of the resonant angle $\theta$. When it librated about $0^{\circ}$, $180^{\circ}$ or other values, we labelled the region QS, HS, or TP$_4$ (or TP$_5$), respectively. In the pale regions, the resonant angle $\theta$ rotates. All types of orbits in co-orbital dynamics are revealed in the DS maps with the boundaries of different domains having been delineated \citep[see also][for more details on the interchange of such regions as the  problem's parameters vary for a different value of $\mu$]{Namouni99,nachrimu99,Nesvorny02,Sidorenko14}. The breakdown of the phase space was organised as follows.

\begin{figure*}[!h]\centering
$\begin{array}{c}
                \includegraphics[width=0.99\textwidth]{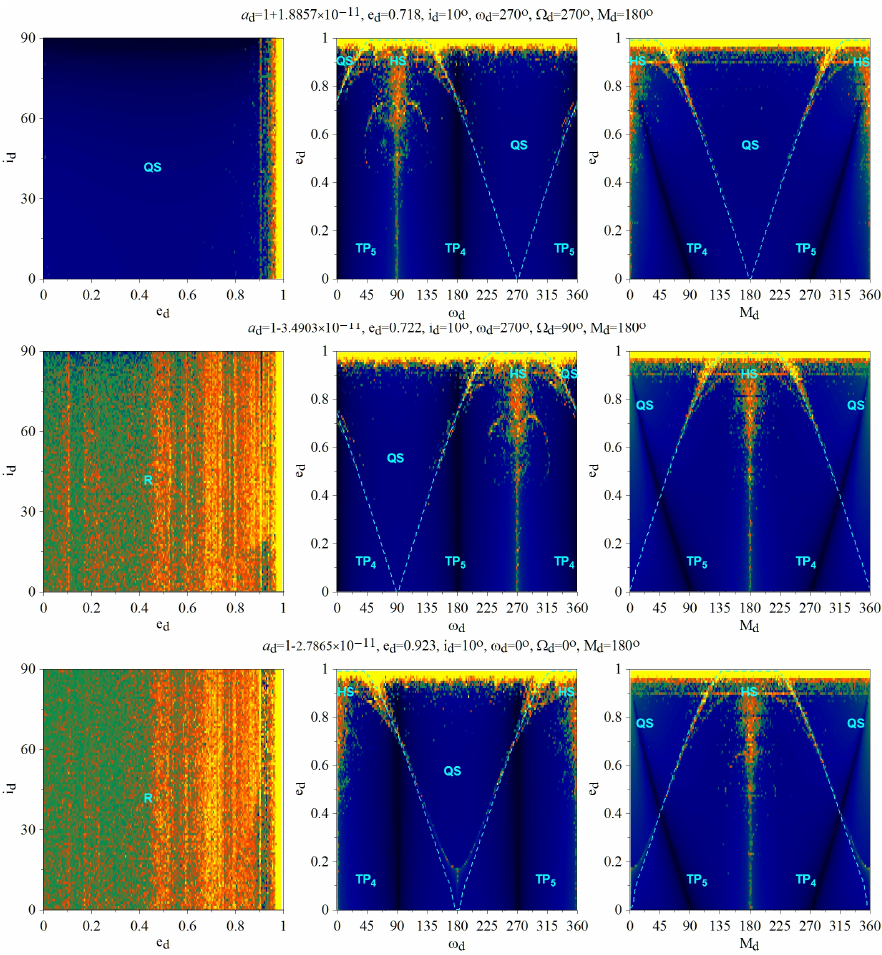}\\
                                                \includegraphics[width=0.2\textwidth]{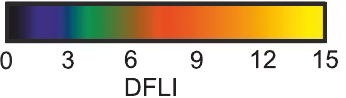}\end{array}$
\caption{DS maps for $\Delta i=10^{\circ}$ guided by a stable periodic orbit of the $F^I$ family (top) and an unstable periodic orbit of the $F^{II'}$ family (middle) and $G^{II'}$ family (bottom). The chosen initial conditions, which remained fixed while some varied for the computation of each grid, are reported above each row (see also Table \ref{tab3}). The dashed cyan curves depict the collisions and close encounters between the bodies. In the areas denoted by QS and HS, $\theta$ librates about 0 and $180^{\circ}$, respectively. TP$_4$ and TP$_5$ stand for the regular domains created by the asymmetric families generated by $L_4$ and $L_5$, where $\theta$ librates accordingly. R denotes areas in which $\theta$ rotates. Dark (pale) colours show regular (chaotic) orbits.}
        \label{2DC_S10}
\end{figure*}

\begin{figure*}[!h]\centering
$\begin{array}{c}
                \includegraphics[width=0.99\textwidth]{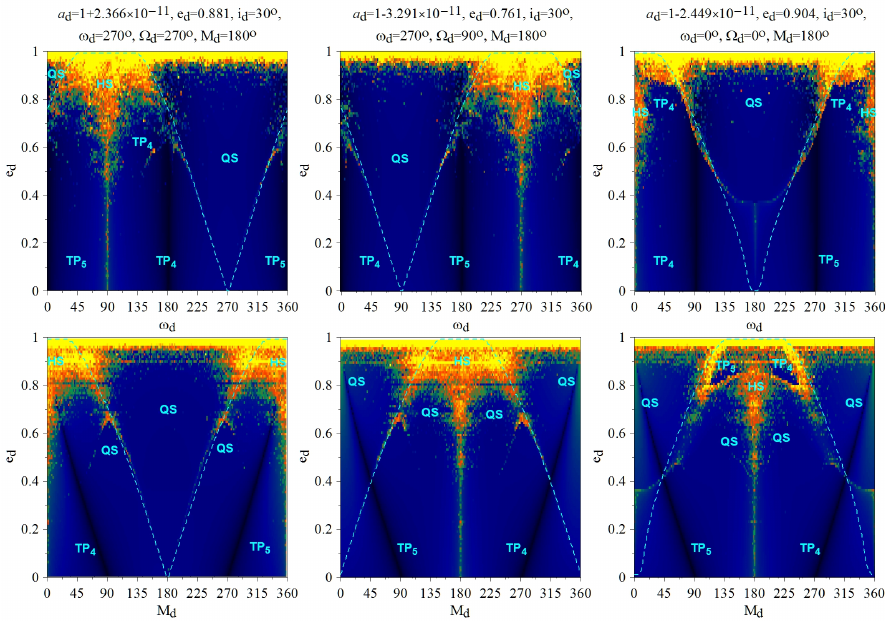}\\
                                                \includegraphics[width=0.2\textwidth]{bar.pdf}\end{array}$
\caption{DS maps for $\Delta i=30^{\circ}$ guided by a stable periodic orbit of the $F^I$family (left) and an unstable periodic orbit of the $F^{II'}$family (middle) and $G^{II'}$family (right). The chosen initial conditions, which remained fixed while some varied for the computation of each grid, are reported above each column (see also Table \ref{tab3}). Presentation as in Fig.~\ref{2DC_S10}.}
        \label{2DC_S30}
\end{figure*}

\begin{figure*}[!h]\centering
$\begin{array}{c}
                \includegraphics[width=0.66\textwidth]{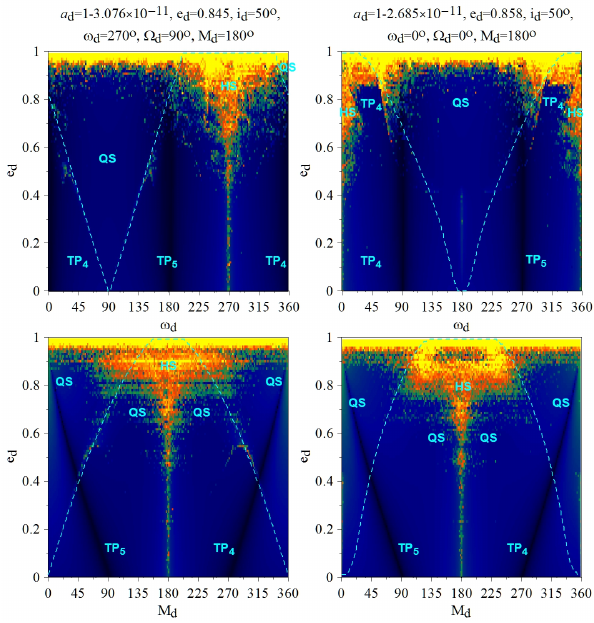}\\
                                                \includegraphics[width=0.2\textwidth]{bar.pdf}\end{array}$
\caption{DS maps for $\Delta i=50^{\circ}$ guided by an unstable periodic orbit of the $F^{II'}$family (left) and $G^{II'}$family (right).  Presentation as in Fig.~\ref{2DC_S30}.}
        \label{2DC_S50}
\end{figure*}

First, we chose three 3D periodic orbits (one from each family, namely $F^I$, $F^{II'}$ and $G^{II'}$) with mutual inclinations, $\Delta i$, equal to $10^{\circ}$, $30^{\circ}$, and $50^{\circ}$ (Figs.~\ref{2DC_S10}-\ref{2DC_S50}). Then, we varied some of the initial conditions to create the grids on the $(e_{\rm d},i_{\rm d})$, $(\omega_{\rm d},e_{\rm d})$, and $(M_{\rm d},e_{\rm d})$ plane, while keeping the rest fixed. The values of the angles are given in Table~\ref{tab3}, and the orbital elements of the selected periodic orbits are also provided in each figure. 

In Fig.~\ref{2DC_S10},we chose three different 3D periodic orbits in which the dust particle had an inclination equal to $10^{\circ}$. In the top row, the periodic orbit that guided the search is stable and belongs to the $F^I$ family with $a_{\rm d}=1+1.887\times 10^{-11}$, $e_{\rm d}=0.718$, and $\Delta\Omega =270^{\circ}$. The areas around the stable 3D orbit are populated with regular orbits, and chaoticity becomes apparent for $e_{\rm d}>0.7$ in the top left panel, where the planar family, $I$, becomes vertically unstable (dashed line in Fig.~\ref{fig3}). In the middle and bottom rows, the periodic orbits that guided the search are unstable and belong to the $F^{II'}$ and $G^{II'}$ families, respectively. The $xz$-symmetric family has initial conditions $a_{\rm d}=1-3.4903\times 10^{-11}$, $e_{\rm d}=0.722$, and $\Delta\Omega =90^{\circ}$, and the $x$-symmetric family has $a_{\rm d}=1-2.7865\times 10^{-11}$, $e_{\rm d}=0.923$, and $\Delta\Omega =0^{\circ}$. The regions around both of them are dominated by irregular orbits. However,  the remaining regular domains QS and TP$_4$ (or TP$_5$) become apparent farther away from this periodic orbit with changing angles.

\begin{figure*}[!h]\centering
$\begin{array}{c}
                \includegraphics[width=0.99\textwidth]{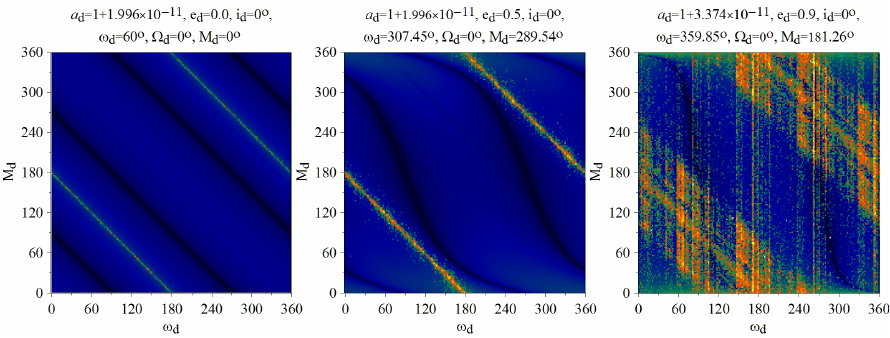}\\
                                                \includegraphics[width=0.2\textwidth]{bar.pdf}\end{array}$
\caption{DS maps on the $(\omega_{\rm d},M_{\rm d})$ plane around asymmetric periodic orbits with $e_{\rm d}=0$ (left), $e_{\rm d}=0.5$ (middle), and $e_{\rm d}=0.9$ (right).}%presented as in Fig.~\ref{2DC_S30}.
        \label{asym_wM}
\end{figure*}

\begin{figure*}[!h]\centering
$\begin{array}{c}
                \includegraphics[width=0.99\textwidth]{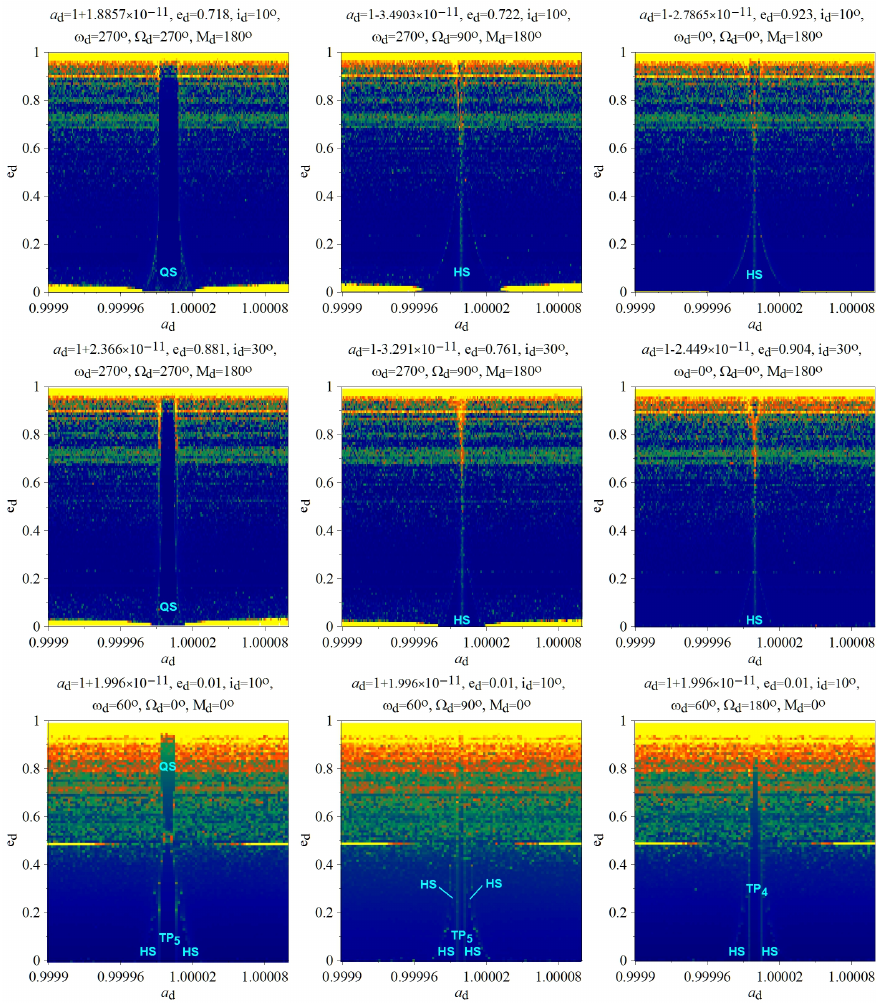}\\
                                                \includegraphics[width=0.2\textwidth]{bar.pdf}\end{array}$
\caption{DS maps on the $(a_{\rm d},e_{\rm d})$ plane guided by symmetric periodic orbits with $\Delta i=10^{\circ}$ (top) and $\Delta i=30^{\circ}$ (middle) and by asymmetric configurations with $\Delta i=10^{\circ}$ (bottom).}% presented as in Fig.~\ref{2DC_S30}
        \label{ae_Di10}
\end{figure*}

For Fig.~\ref{2DC_S30}, we performed the same search, but chose three different 3D periodic orbits in which the dust particle had an inclination equal to $30^{\circ}$. We only present grids on the $(\omega_{\rm d},e_{\rm d})$ and $(M_{\rm d},e_{\rm d})$ plane, as the variation in the $(e_{\rm d},i_{\rm d})$ plane was qualitatively similar with that presented in the left column of Fig.~\ref{2DC_S10}. In the left column of Fig.~\ref{2DC_S30}, the stable 3D periodic orbit has $a_{\rm d}=1+2.366\times 10^{-11}$, $e_{\rm d}=0.881$, and $\Delta\Omega =270^{\circ}$. This 3D periodic orbit is shown in panel a of Fig. \ref{QSorbits3D}. In the middle column, the 3D unstable orbit has $a_{\rm d}=1-3.291\times 10^{-11}$, $e_{\rm d}=0.761$, and $\Delta \Omega =90^{\circ}$, and in the right column, the initial conditions are $a_{\rm d}=1-2.449\times 10^{-11}$, $e_{\rm d}=0.904$, and $\Delta\Omega =0^{\circ}$. 

For Fig.~\ref{2DC_S50}, we performed the same search, but chose two unstable 3D periodic orbits in which the dust particle had an inclination equal to $50^{\circ}$ with $a_{\rm d}=1-3.076\times 10^{-11}$, $e_{\rm d}=0.845$, and $\Delta\Omega =90^{\circ}$ (left column) and $a_{\rm d}=1-2.685\times 10^{-11}$, $e_{\rm d}=0.858$, and $\Delta\Omega =0^{\circ}$ (right column). The latter 3D periodic orbit is shown in panel b of Fig. \ref{QSorbits3D}. The chaotic domains created around the 3D HS orbits span greater areas with increasing inclination and reach even lower eccentricity, $e_{\rm d}$, values for the dust particle.

In Fig.~\ref{asym_wM} we created DS maps on the $(\omega_{\rm d},M_{\rm d})$ plane around three 2D asymmetric periodic orbits with $e_{\rm d}=0$ (left panel), $e_{\rm d}=0.5$ (middle panel), and $e_{\rm d}=0.9$ (right panel). The regular orbits dominate as the angles vary as long as $e_{\rm d}$ does not approximate the bifurcation point on the HS orbit, that is, at 0.917.

In Fig.~\ref{ae_Di10} we varied the semi-major axis and the eccentricity of the dust particle on the $(a_{\rm d},e_{\rm d})$ plane. In the top and middle rows, we used the same six periodic orbits as in Figs.~\ref{2DC_S10} and \ref{2DC_S30} with $\Delta i=10^{\circ}$ and $\Delta i=30^{\circ}$, respectively. Therefore, islands of stability around the QS orbits were created in the DS maps. In the bottom row, we explored the spatial neighbourhood of the planar vertically stable asymmetric TP orbit close to $L_4$ (or $L_5$) for $\Delta i=10^{\circ}$ and three different nodal differences, namely $\Delta\Omega=0^{\circ}$, $90^{\circ}$ and $180^{\circ}$. These numerical results and the boundaries of the domains, namely the libration of the resonant angle, agree with those presented by \citet{Nesvorny02}.

\section{Setup of the $N$-body simulations}\label{nbody}

After computing periodic orbits and creating DS maps, we performed $N$-body simulations. In order to accurately compare these simulations to the DS maps, care must be taken with regard to rescaling \citep{ave16,ave19}.

The setup of the $N$-body simulations requires that the data extracted from the families of periodic orbits and the DS maps are transformed into real units. In this case, the results presented here can be applied to different planetary system configurations. For instance, different units of time, masses, and distances can be used. We needed to set a normalisation with respect to the masses assumed for the computation of the periodic orbits in our case. The details of the invariance of the equations of motion we used to compute the periodic orbits and the appropriate scaling factors were explained by \citet{ave16} for the 2D-CRTBP and by \citet{ave19} for the 2D-ERTBP and the two 3D-RTBPs.

We performed a scaling in our simulations with regard to the semi-major axis of the asteroid and the dust particle, so that 
 \begin{equation}
a_i^{\rm (N)} = a_i \zeta^{1/3},
\end{equation}
 where $\zeta \equiv \frac{m_{\rm WD} + m_{\rm a}}{1m_{\sun}}$ is the scaling factor, and $\rm N$ represents the scaled orbital elements used in the $N$-body simulations. Therefore, the  values of the semi-major axes that were imported to the $N$-body integrator were multiplied by the factor 0.843432665311676. Time and the remainder of the orbital elements (e.g. the eccentricities and angles) remained the same. 

For all of our $N$-body simulations, we used the {\tt IAS15} integrator in the {\tt REBOUND} simulation package \citep{rein2012,rein2015}. This integrator uses an adapative time step and conserves energy sufficiently well for our purposes. In order to model a realistically observable timescale, we performed each simulation over 10 years, corresponding to roughly 20,000 orbits of the minor planet. We output data snapshots every 0.05 years, but monitored if and when collisions occurred for every time step. Although our test particles were point masses, our minor planets were not: Their radius was 200~km.

Limiting the integration timescale to 10 years also prevented longer-term forces on the dust particles from becoming important. Several shorter-term forces might also act on the particles. We discuss how these can be contextualized within the outputs of our simulations in Sect. \ref{dis}. These forces typically have a greater effect the smaller the orbital pericentre. In this context, the most consequential results are those with $e_{\rm d} \approx 0.0$, whereas those with more limited applicability have $e_{\rm d} \gtrsim 0.8$. Nevertheless, we sampled $e_{\rm d}$ values up to 0.9 in many cases to provide a direct comparison with the DS maps.

\section{Results of the $N$-body simulations}\label{sims}

We had a large number of parameters to select in order to report our results. We chose a parameter that has a direct observable link: the maximum variation in the orbital period of the dust particle. The orbital period is directly measured and is independent of other parameters. Transit photometry in these WD systems can be measured on the scale of seconds \citep{gan16,farihi22,agd24}.

We wished to determine the extent of this orbital period variation and how it relates, if it does, to periodic orbits and the structure seen in the DS maps. Because of the computational expense of integrating over a grid of initial conditions with an accurate time-variable integrator such as {\tt IAS15}, we selected a few parts of a few parameter grids over which to perform simulations. These are outlined below. In all cases, each parameter was sampled uniformly.

For some parameter combinations, some systems become unstable. These instabilities are manifested entirely through collisions between a dust particle and the asteroid. In our simulations, no dust particles escape the system and no dust particles collide with the WD. Unstable systems are indicated by red crosses in the plots.

Within the set of six orbital parameters, the only parameter that determines the location of the dust particle rather than setting the shape of its orbit is $M_{\rm d}$. In some cases, in order to provide a direct comparison to the DS maps, we adopted the value of $M_{\rm d}$ that corresponds to a particular family of periodic orbit, or DS map. However, in some other cases, we sampled  a range of $M_{\rm d}$ values in order to determine the maximum period deviation independent of $M_{\rm d}$, and we report only those that generated the maximum orbital period deviation.

\subsection{Variation in $e_{\rm d}$ versus $a_{\rm d}$} 

\begin{figure*}\centering
$\begin{array}{cc}
\includegraphics[width=0.45\textwidth]{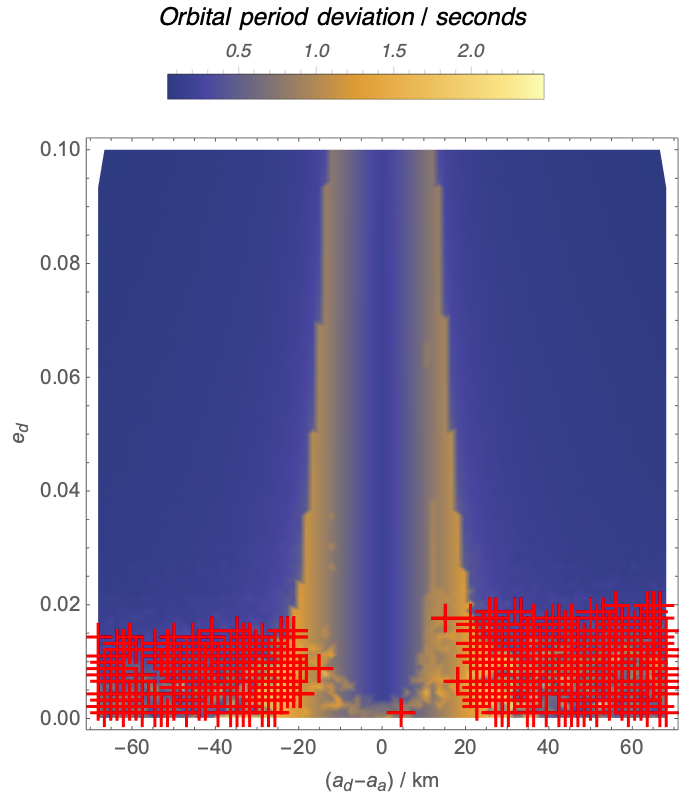}&\includegraphics[width=0.45\textwidth]{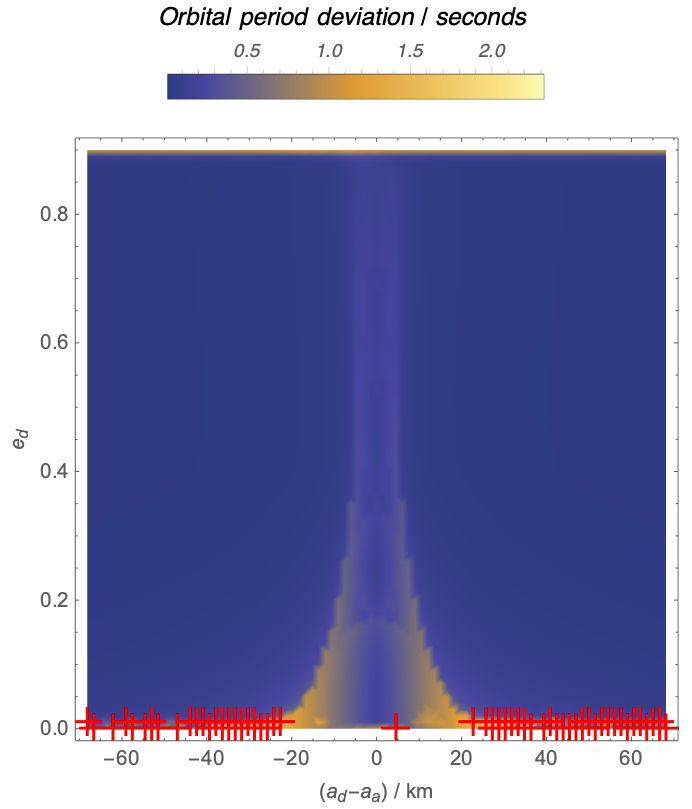}\\
\includegraphics[width=0.5\textwidth]{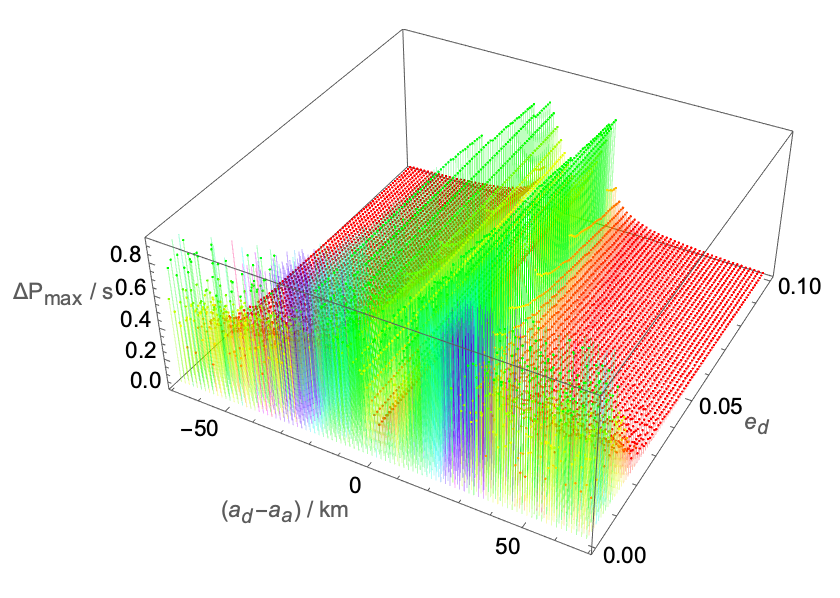}&\includegraphics[width=0.5\textwidth]{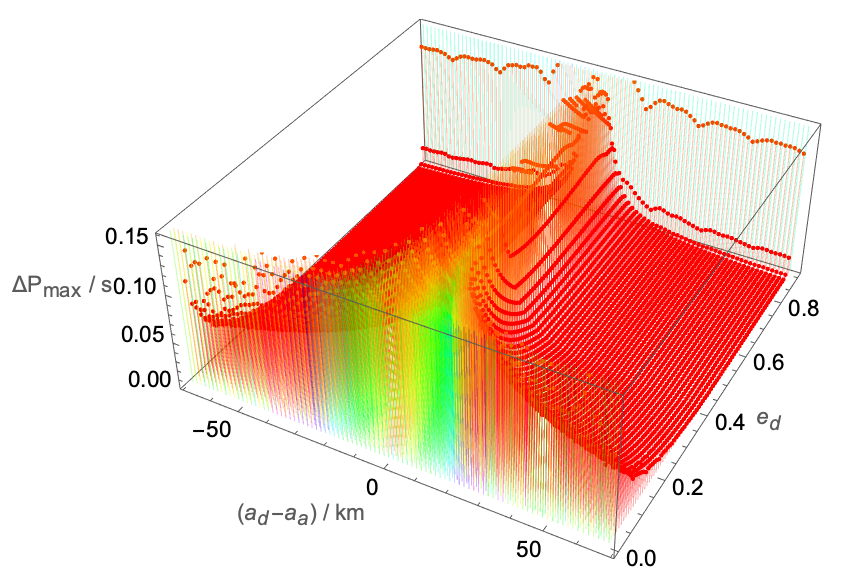}  \end{array}$
\caption{Simulations on the $(\Delta a, e_{\rm d})$ plane (top row) and in the $(\Delta a, e_{\rm d}, \Delta P_{\rm max})$ space (bottom row) guided by the 3D QS periodic orbit with $\Delta \Omega=270^{\circ}$ and $\Delta i=10^{\circ}$ that formulated the respective DS map in Fig. \ref{ae_Di10} (top left panel). Here, $a_{\rm d} = a_{\rm a} \mp 68.135~{\rm km}$, and $M_{\rm d}$ is uniformly sampled and incorporates orbits around the 3D QS periodic orbit, and the left column represents a magnification for $e_{\rm d} < 0.1$. Unstable systems are indicated by red crosses on the $(\Delta a, e_{\rm d})$ plane.}  
        \label{PlotConFig12TopLeft}
\end{figure*}

Fig.~\ref{PlotConFig12TopLeft} shows a $91 \times 91$ grid of $a_{\rm d} = a_{\rm a} \mp 68.135~{\rm km}$ and $e_{\rm d} = 0.0001-0.1000$ (left column) and $e_{\rm d} = 0.0001-0.9000$ (right column). For each combination of $(a_{\rm d}, e_{\rm d})$, we performed 18 simulations with $M_{\rm d}$ uniformly sampled across its entire range. The maximum orbital period deviation occurs along the boundaries of the island of stability that is populated by low-eccentricity orbits with $\Delta \Omega=270^{\circ}$ and $\Delta i=10^{\circ}$.

\begin{figure}\centering
$\begin{array}{c}
\includegraphics[width=0.45\textwidth]{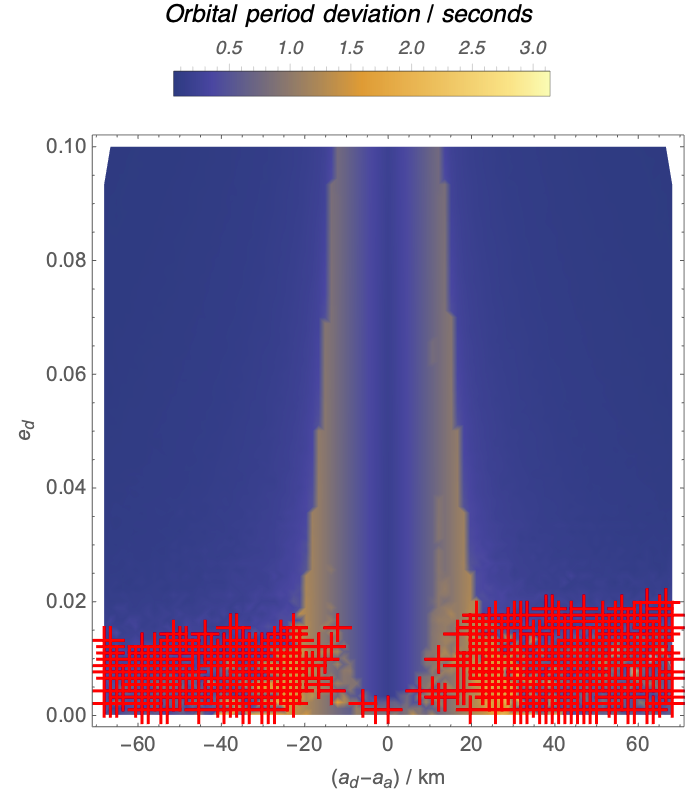}\\
\includegraphics[width=0.5\textwidth]{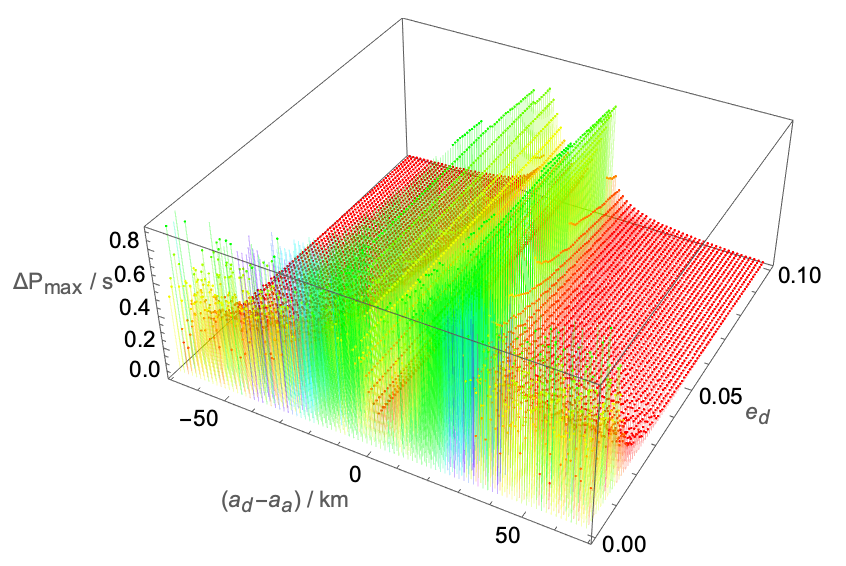}  \end{array}$
\caption{Presentation as in Fig. \ref{PlotConFig12TopLeft}, but guided by the 3D HS periodic orbit with $\Delta \Omega=90^{\circ}$ and $\Delta i=10^{\circ}$ that formulated the respective DS map in Fig. \ref{ae_Di10} (top middle panel). Unstable systems are indicated by red crosses in the $(\Delta a, e_{\rm d})$ plane.} 
        \label{PlotConFig12TopMiddle}
\end{figure}

A similar behaviour is shown in Fig. \ref{PlotConFig12TopMiddle}, but guided by the 3D HS periodic orbit with $\Delta \Omega=90^{\circ}$ and $\Delta i=10^{\circ}$. Again, the maximum orbital period deviation takes place along the boundaries of the island where $\Delta \Omega=90^{\circ}$. 

\begin{figure}\centering
$\begin{array}{c}
\includegraphics[width=0.45\textwidth]{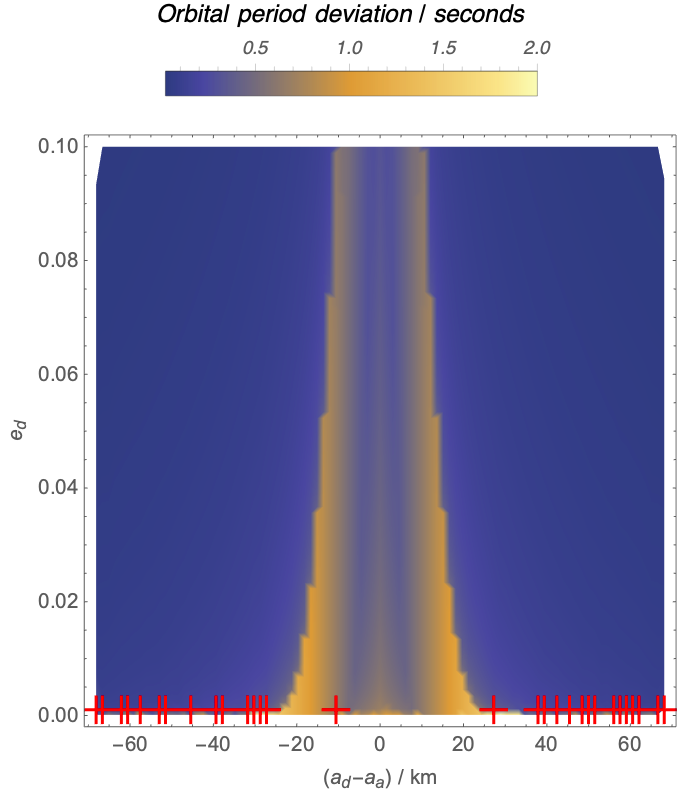}\\
\includegraphics[width=0.5\textwidth]{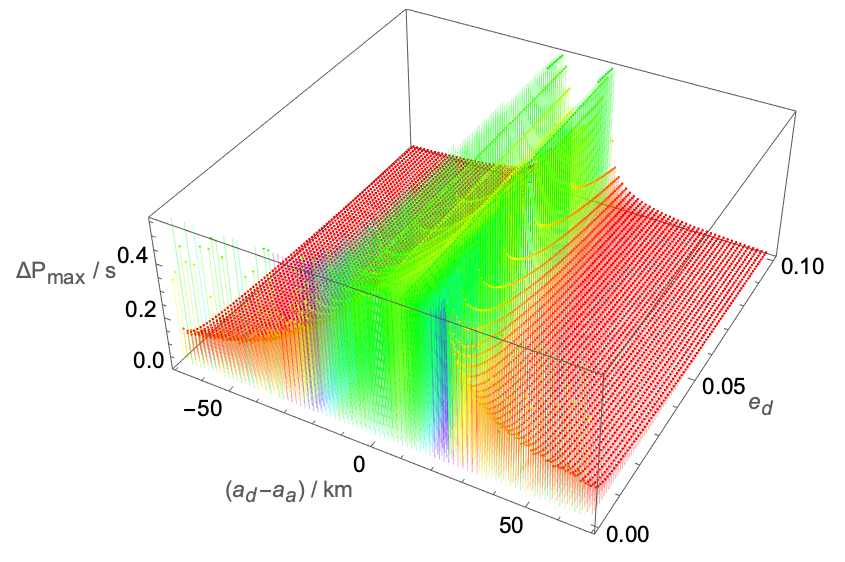}  \end{array}$
\caption{Presentation as in Fig. \ref{PlotConFig12TopLeft}, but guided by the 3D HS periodic orbit with $\Delta \Omega=0^{\circ}$ and $\Delta i=10^{\circ}$ that formulated the respective DS map in Fig. \ref{ae_Di10} (top right panel). Unstable systems are indicated by red crosses on the $(\Delta a, e_{\rm d})$ plane.}.  
        \label{PlotConFig12TopRight}
\end{figure}

Fig. \ref{PlotConFig12TopRight} shows a grid guided by the 3D HS periodic orbit with $\Delta \Omega=0^{\circ}$ and $\Delta i=10^{\circ}$. Again, the maximum orbital period deviation occurs along the boundaries of the respective island of stability. 

\begin{figure}\centering
$\begin{array}{c}
\includegraphics[width=0.45\textwidth]{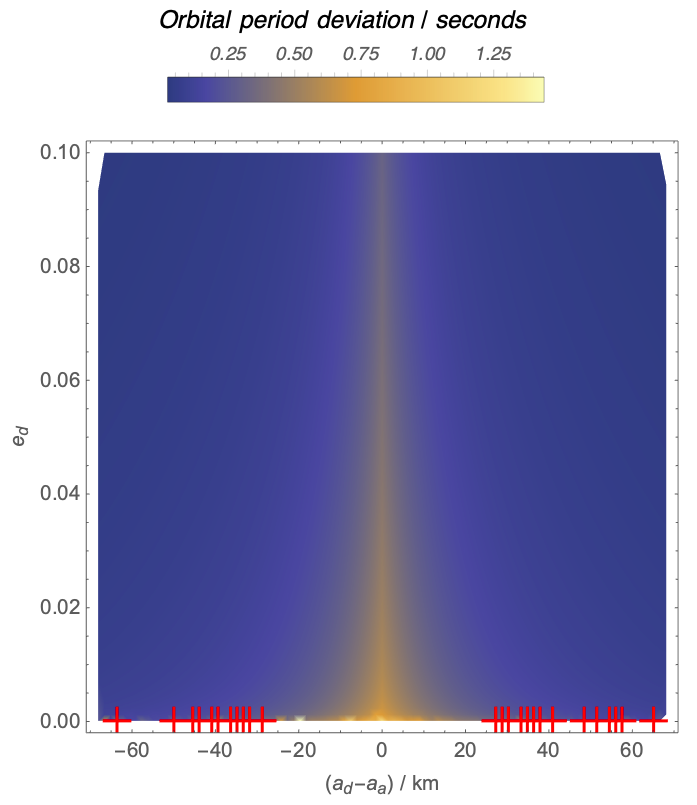}\\
\includegraphics[width=0.5\textwidth]{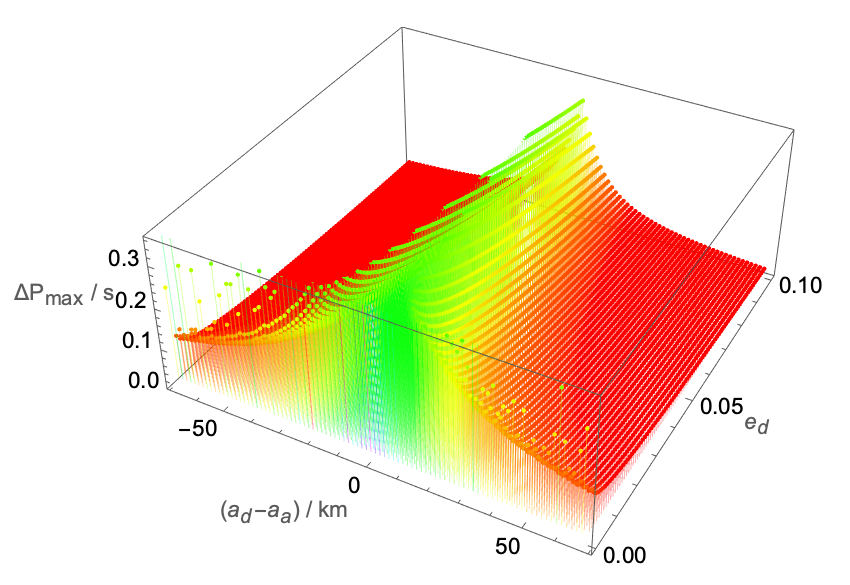}  \end{array}$
\caption{Presentation as in Fig. \ref{PlotConFig12TopLeft}, but guided by a TP periodic orbit with $\omega_{\rm d}=60^{\circ}$ and $\Delta i=10^{\circ}$ that formulated the respective DS map in Fig. \ref{ae_Di10} (bottom left panel). $M_{\rm d}$ is not uniformly sampled, but equals $0^{\circ}$. Unstable systems are indicated by red crosses in the $(\Delta a, e_{\rm d})$ plane.} 
        \label{PlotConFig12BottomLeft}
\end{figure}

Fig. \ref{PlotConFig12BottomLeft} shows a grid guided by a TP periodic orbit with $\omega_{\rm d}=60^{\circ}$ and $\Delta i=10^{\circ}$. In contrast to the previous simulations, we performed a single simulation with $M_{\rm d} = 0^{\circ}$. Hence, the value of the maximum orbital period deviation observed in the region populated by TP orbits was lower. 

\subsection{Variation in $e_{\rm d}$ versus $\omega_{\rm d}$}

\begin{figure*}\centering
$\begin{array}{cc}
\includegraphics[width=0.45\textwidth]{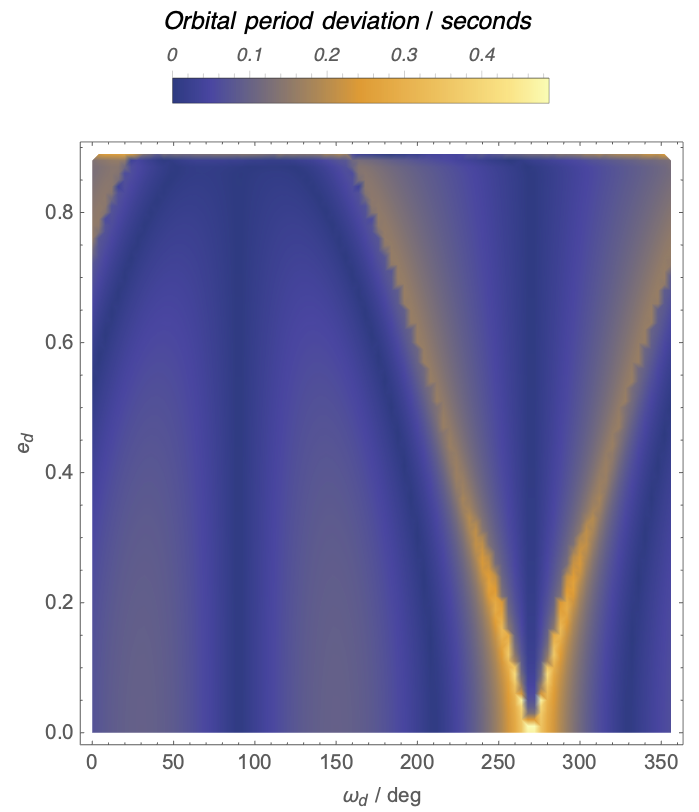}&\includegraphics[width=0.45\textwidth]{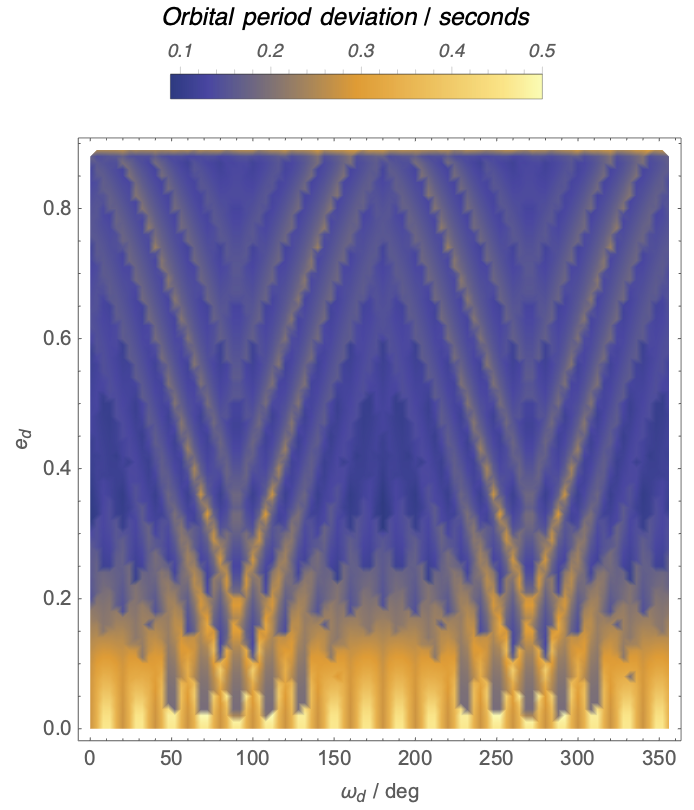}\\
\includegraphics[width=0.5\textwidth]{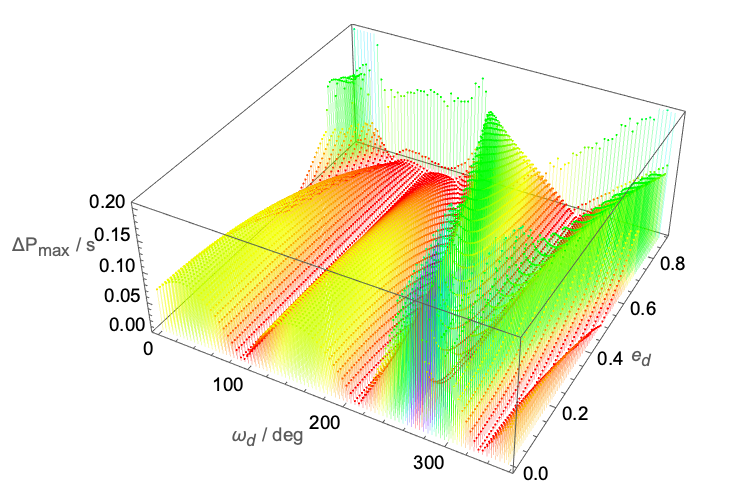}&\includegraphics[width=0.5\textwidth]{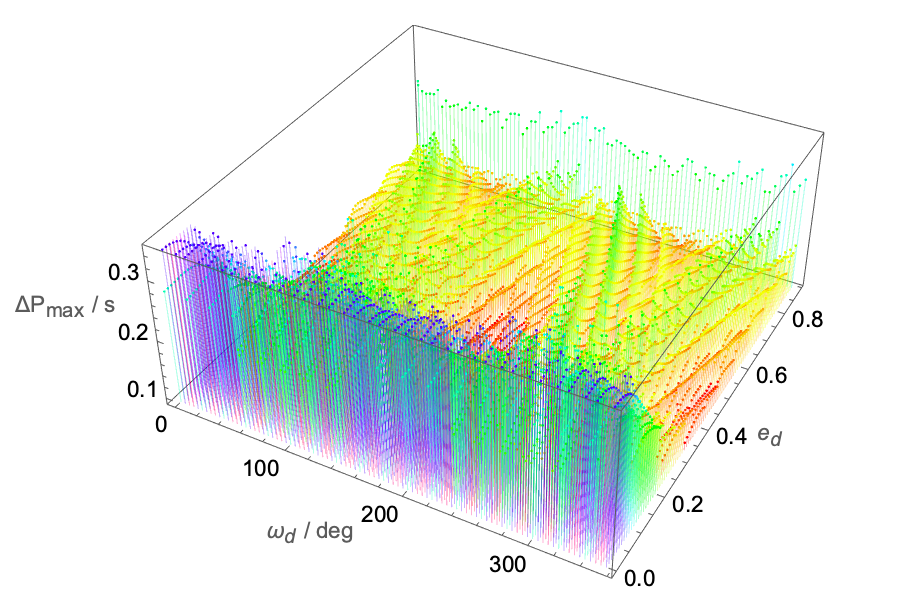}  \end{array}$
\caption{Simulations on the $(\omega_{\rm d}, e_{\rm d})$ plane (top row) and in the $(\omega_{\rm d}, e_{\rm d}, \Delta P_{\rm max})$ space (bottom row) guided by the 3D QS periodic orbit with $\Delta \Omega=270^{\circ}$ and $\Delta i=10^{\circ}$ that formulated the respective DS map in Fig. \ref{2DC_S10} (top middle panel). In the left column, $M_{\rm d}= 180^{\circ}$, while it is uniformly sampled and incorporates orbits around the 3D QS periodic orbit on the right.}  
        \label{PlotConFig8Top}
\end{figure*}

Fig.~\ref{PlotConFig8Top} shows a $91 \times 91$ grid of $\omega_{\rm d} = 0^{\circ}-360^{\circ}$ and $e_{\rm d} = 0.0001-0.9000$ guided by the 3D QS periodic orbit with $\Delta \Omega=270^{\circ}$ and $\Delta i=10^{\circ}$. For each combination of $(\omega_{\rm d}, e_{\rm d})$, we performed a single simulation with $M_{\rm d} = 180^{\circ}$ (left column) and 18 simulations with $M_{\rm d}$ uniformly sampled across its entire range (right column).  When $M_{\rm d} = 180^{\circ}$, the correspondence between the DS map and the $N$-body simulations is excellent, although the maximum orbital period deviation observed was negligible in this exploration. This deviation did not increase even when a uniform sample was adopted for $M_{\rm d}$.

\subsection{Variation in $i_{\rm d}$ versus $e_{\rm d}$}

\begin{figure}\centering
$\begin{array}{c}
\includegraphics[width=0.45\textwidth]{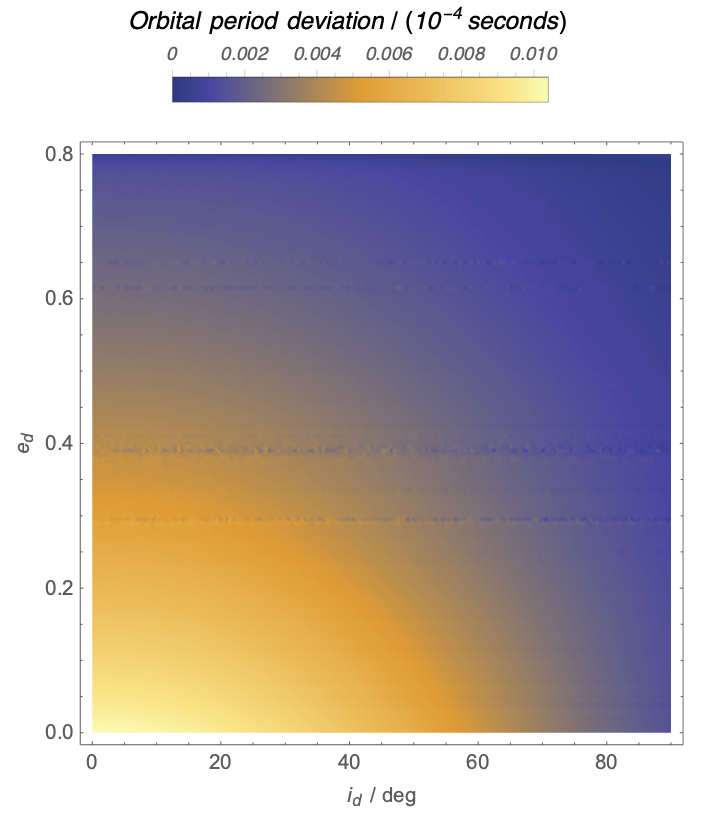}\\
\includegraphics[width=0.5\textwidth]{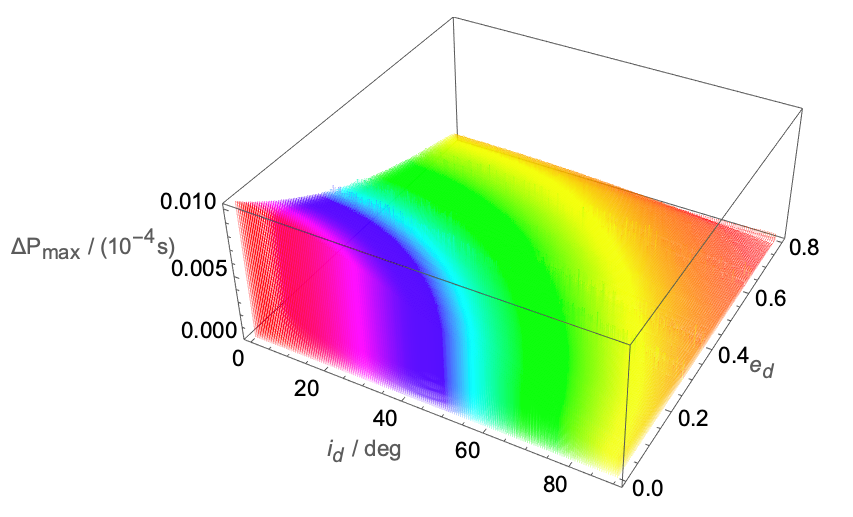}  \end{array}$
\caption{Simulations on the $(i_{\rm d}, e_{\rm d})$ plane (top row) and in the $(i_{\rm d}, e_{\rm d}, \Delta P_{\rm max})$ space (bottom row) guided by a HS periodic orbit with $\Omega_{\rm d}=90^{\circ}$ and $\Delta i=50^{\circ}$ that formulated the respective DS maps in Fig. \ref{2DC_S50} (left column). $M_{\rm d}$ is not uniformly sampled, but instead equals $180^{\circ}$ in each simulation.} 
        \label{PlotConFig10Left}
\end{figure}

Fig. \ref{PlotConFig10Left} shows a $181 \times 181$ grid of $e_{\rm d} = 0.0001-0.8000$ and $i_{\rm d} = 0^{\circ}-90^{\circ}$ guided by the HS periodic orbit with $\Omega_{\rm d}=90^{\circ}$ and $\Delta i=50^{\circ}$, and we performed a single simulation with $M_{\rm d} = 180^{\circ}$. The  maximum orbital period deviation observed around the 3D HS orbits was negligible, as the semi-major axis of the dust was fixed and equal to that of the periodic orbit. 

\section{Discussion}\label{dis}

Accurately modelling WD debris discs requires incorporating a vast variety of physical processes, including gas-dust interactions, gas drag, aeolian erosion, Ohmic heating, Lorentz drift, and external perturbations. We have isolated the effect of an important and fundamental force that acts on all the discs: the effect of gravity, and in the likely co-orbital scenario with a disrupted asteroid.

Nevertheless, there are some key physics that we highlight below and relate to our results.

\subsection{General relativity}

Similar to Newtonian gravity, general relativity affects all objects orbiting WDs. Furthermore, the effect is greater than that of most known main-sequence extrasolar systems because WD debris discs are so compact. 

Over the course of a single orbit, general relativity changes each orbital element \citep{li12,verGR14}. However, these changes all average out to zero, except for the change in the argument of pericentre. Hence, for circular orbits, the effect of general relativity is negligible. For non-circular orbits, in WD debris discs, $\omega_{\rm d}$ cycles across its entire range in a time of approximately
% just decades (the exact time depends on $e_{\rm d}$ and $M_{\star}$).

\begin{equation}
107~{\rm yr} \left( \frac{M_{\star}}{0.60~M_{\odot}} \right)^{-3/2}
             \left( \frac{a_{\rm d}}{1.00R_{\odot}}  \right)^{5/2}
             \left( 1-e_{\rm d}^2 \right).
\end{equation}

As a result, we might imagine a dust particle traversing a horizontal path across the plots in Figs. \ref{2DC_S10}-\ref{asym_wM}. However, because the maximum period deviation in the entire phase space is less than a few seconds, this traversal would probably not be observable, regardless of the manner in which it takes place.

\subsection{Sublimation}

%A boulder, pebble or dust particle may be treated similarly to a gas particle in the context of a gravitational three-body problem with a WD and a minor planet. In this respect, whether or not a dust particle sublimates should not affect our results. 

The creation of gas through dust sublimation generates gas-dust interactions, which are still poorly understood and are difficult to model analytically. Hence, our results are best applied to dust alone.

The location in which dust sublimates depends on the temperature of the WD \citep{Steckloff2021}, which in turn depends on how long it has been a WD. This time span is commonly known as its cooling age. WDs have been observed with cooling ages of up to $\approx $10 Gyr \citep{elms22}. Old WDs hence allow the orbital pericentre of a dust particle to be far inwards of $1R_{\odot}$ without sublimating.

\begin{figure}
\begin{center}
\includegraphics[width=0.5\textwidth]{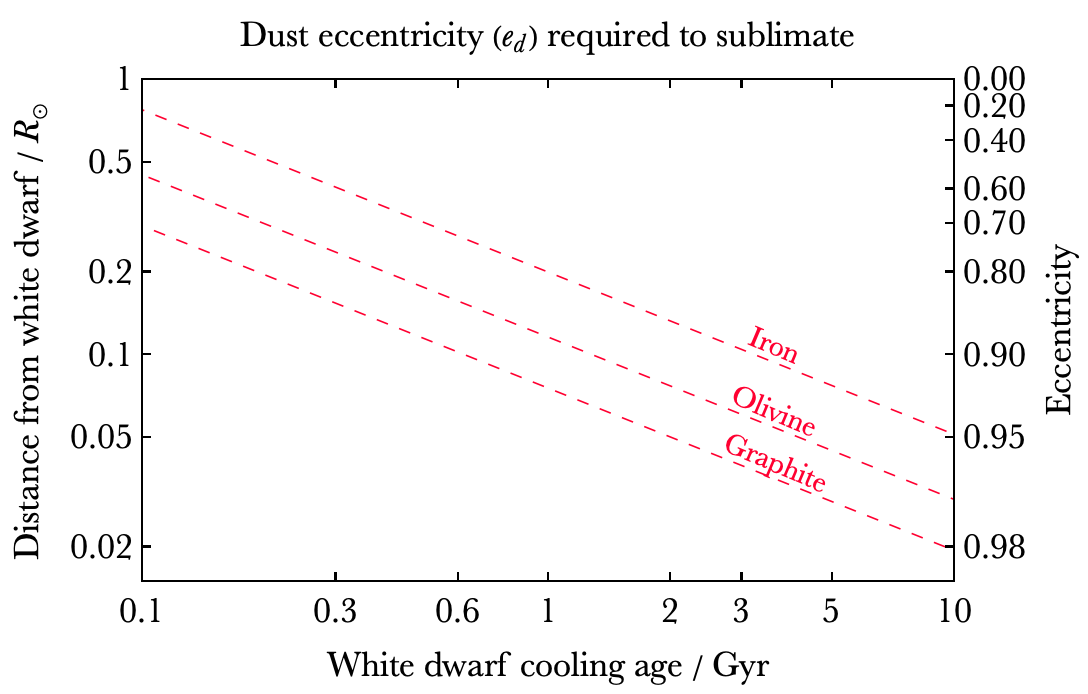}
\end{center}
\caption{Maximum eccentricity values of the dust particle (right $y$-axis), consisting of three different materials, before the particles are sublimated, as a function of the WD cooling age. Stronger materials may reach $e_{\rm d}\approx 0.8-0.9$ without sublimating when the WD is several billion years old.} 
        \label{sublim}
\end{figure}

In order to quantify this effect, we computed the maximum value of $e_{\rm d}$ allowed for a dust particle to survive as dust before it is sublimated as a function of WD cooling age. To do this, we used the same prescription as in \cite{vbz22}, and we report our results in Fig.~\ref{sublim}. This prescription involves combining equations for the blackbody radiation of the WD, the luminosity of a WD as a function of cooling age, and a relation between effective temperature, distance, and sublimation temperatures of three different materials. The plot considers three different potential compositions of the dust and shows that in many cases, values of $e_{\rm d}\approx 0.8-0.9$ may realistically be adopted.

\subsection{Poynting-Robertson drag}

Even when a dust particle is not sublimated, the radiation from the WD will drag the particle inwards by Poynting-Robertson drag. This drag occurs for all dust particles and for pebbles, boulders, and any other objects that would be considered as test particles in this study. Because the speed at which Poynting-Robertson drag acts strongly depends on the size of the particle, the relevance of this effect in our study in turn strongly depends on our assumption of the size of the test particle.

We can estimate the inward drift of a particle through the classic \citet{Wyatt1950} formula for the semi-major axis variation due to Poynting-Robertson drag,

\begin{equation}
\frac{da}{dt} \approx - \frac{3L_{\star}\left(2 + 3 e_{\rm d}^2\right)}{16 \pi c^2 R_{\rm d} \rho_{\rm d} a_{\rm d} \left(1 - e_{\rm d}^2 \right)^{3/2}},
\end{equation}

\noindent{}and we assumed that the particle density is $\rho_{\rm d} = 2$~g/cm$^3$.

\begin{figure}
\begin{center}
\includegraphics[width=0.5\textwidth]{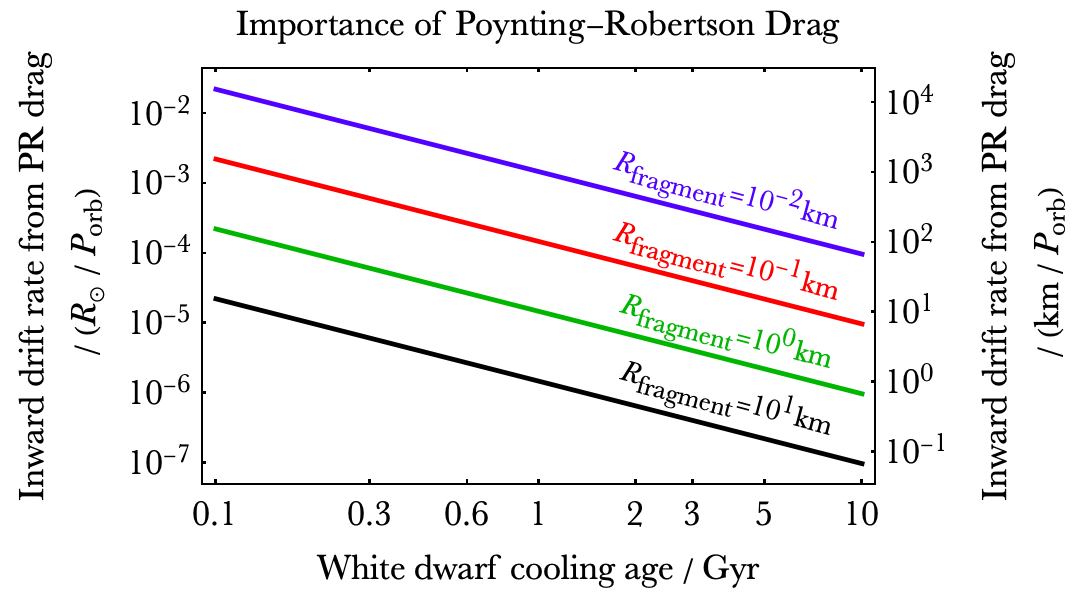}
\end{center}
\caption{Inward drift rate from Poynting-Robertson drag as a function of the WD cooling age. Even the largest boulders may be dragged enough to trigger gravitational close encounters with the co-orbital asteroid.}
        \label{PRDrag}
\end{figure}

Figure~\ref{PRDrag} illustrates the drag rate for four different particle sizes that are large enough to be more appropriately classed as boulders. Over a 10 yr time span, even the largest boulders would be dragged inwards to a greater extent than their orbit would be gravitationally shifted due to a minor planet in a stable co-orbital configuration. We therefore suggest that observational evidence of a period shift is much more likely to arise from Poynting-Robertson drag than from co-orbital dynamics.

We can further deduce that for $e_{\rm d} \approx 0$, based on the x-axes of the $N$-body simulation plots for Figs. \ref{PlotConFig12TopLeft}-\ref{PlotConFig12BottomLeft}, a dust particle that starts in a stable co-orbital configuration with a minor planet can be dragged inwards by $\approx$20-60~km into a region of gravitational instability. In this case, gravity might ultimately generate a collision between the particle and minor planet, even though the trigger was radiative drag.

\section{Conclusions}\label{con}

Motivated by asteroids and dust in co-orbital configurations around WDs, we computed the families of symmetric and asymmetric periodic orbits in the 2D and 3D CRTBP for the WD--asteroid--dust particle configuration with $\mu\approx 10^{-11}$ in the 1:1 MMR. The 3D families with unstable periodic orbits presented here are novel. In this configuration, QS orbits were additionally found as $e_{\rm d}\rightarrow 0$. Nonetheless, we found that the planar families exhibited qualitatively similar attributes with those computed for $\mu=0.001$ by \citet{Pousse17} and \citet{va18}, that is, their linear stability and their bifurcation points, even though we approached the unperturbed case. 

We methodically explored the phase space by choosing specific 3D symmetric and 2D asymmetric periodic orbits and monitored the libration of the resonant angle. All types of orbits in co-orbital dynamics were revealed via the DS maps, in which the boundaries of different domains were delineated.

All types of orbits were simulated over a timescale of 10 yr with $N$-body simulations, namely QS, HS, and TP orbits in the 1:1 MMR. In all cases, an observable orbital period deviation was exhibited in the low-eccentricity regime $e_{\rm d}<0.1$ within each domain and nodal difference, while the maximum (about 3 seconds) was yielded when $\Delta \Omega=90^{\circ}$ while drifting away from the periodic orbit ($a_{\rm d}\neq 1$) and being almost at the boundaries of the stability domain. 

Before circularising, an asteroid approaches a WD on an eccentric orbit. Hence, a future useful extension of this work may investigate the three-body problem with an eccentric asteroid at a similarly extreme mass ratio. Even so, whether circular or eccentric, a companion of a WD with transiting dust may have another value of $\mu$ when a planetesimal is considered. The 2D and 3D periodic orbits play a crucial role in unravelling and determining possible regimes with expected transits \citep[see e.g. the stable domains in the DS maps in][for possible detection of terrestrial planets around single-planet systems, i.e. the case of 2D and 3D RTBPs with $\mu=0.001$]{kiaasl,spis,spa}\footnote{Likewise for possible detections in the GTBP, \citep[see e.g.][]{av16,kiaExpl,Antoniadou2022}}. How large $\mu$ needs to be in order for the co-orbital dynamics to become dominant, that is, with an orbital period variation about 3D periodic orbits, namely a detection, remains to be explored.

\begin{acknowledgements} We thank the anonymous reviewer for their astute comments and careful reading which improved the manuscript. The work of KIA was supported by the University of Padua under Grant No. BIRD232319. Results presented in this work have been produced using the Aristotle University of Thessaloniki (AUTh) High Performance Computing Infrastructure and Resources. \end{acknowledgements}

\bibliographystyle{aa}
\bibliography{11}

\clearpage

%\begin{appendix}
%
%\section{Section}\label{appsec}
%
%\end{appendix}

\end{document}